\documentclass[utf8]{FrontiersinHarvard} 

\usepackage{url,hyperref,lineno,microtype,subcaption}
\usepackage[onehalfspacing]{setspace}

\usepackage{psfrag, amsmath, amssymb, amscd, amsfonts, latexsym, epsf, graphicx}
\def\bbbr{{\mathbb R}}
\newcommand{\orth}{\bot}

\def\keyFont{\fontsize{8}{11}\helveticabold }
\def\firstAuthorLast{Lindeberg} 
\def\Authors{Tony Lindeberg\,$^{1,*}$}
\def\Address{$^{1}$Computational Brain Science Lab, Division of Computational Science and Technology, KTH Royal Institute of Technology,  Stockholm, Sweden}
\def\corrAuthor{Tony Lindeberg}

\def\corrEmail{tony@kth.se}

\begin{document}
\onecolumn
\firstpage{1}

\title{Covariance properties under natural image transformations for the generalized Gaussian derivative model for visual receptive fields} 

\author[\firstAuthorLast ]{\Authors} 
\address{} 
\correspondance{} 

\extraAuth{}

\maketitle

\begin{abstract}

  \section{}
  The property of covariance, also referred to as equivariance, means
  that an image operator is well-behaved under image transformations,
  in the sense that the result of applying the image operator to a
  transformed input image gives essentially a similar result as applying the same
  image transformation to the output of applying the image operator to the
  original image. This paper presents a theory of
  geometric covariance properties in vision, developed for a generalized Gaussian
  derivative model of receptive fields in the primary visual cortex
  and the lateral geniculate nucleus, which, in turn, enable
   geometric invariance properties at higher levels in the visual
  hierarchy.
  

It is shown how the studied generalized Gaussian derivative model for
visual receptive fields obeys true covariance properties under spatial
scaling transformations, spatial affine transformations, Galilean
transformations and temporal scaling transformations. These
covariance properties imply that a
vision system, based on image and video measurements in terms of the
receptive fields according to the generalized Gaussian derivative model,
can, to first order of approximation, handle the image and video
deformations between multiple views of objects delimited by smooth
surfaces, as well as between multiple views of spatio-temporal events,
under varying relative motions between the objects and events in the
world and the observer.

We conclude by describing implications of the presented theory for
biological vision, regarding connections between the variabilities of
the shapes of biological visual receptive fields
and the variabilities of spatial and spatio-temporal image structures
under natural image transformations.
Specifically, we formulate experimentally testable biological
hypotheses as well as needs for measuring population statistics of
receptive field characteristics, originating from predictions from the presented theory,
concerning the extent to which the shapes of the biological receptive fields
in the primary visual cortex span the variabilities of spatial and
spatio-temporal image structures induced by natural
image transformations, based on geometric covariance properties.

\tiny
 \keyFont{ \section{Keywords:} receptive field, image transformations,
   scale covariance, affine covariance, Galilean covariance,
   primary visual cortex, vision, theoretical neuroscience} 
\end{abstract}

\section{Introduction}

\begin{figure}[hbtp]
  \begin{center}
    \begin{tabular}{ccc}
      \includegraphics[width=0.30\textwidth]{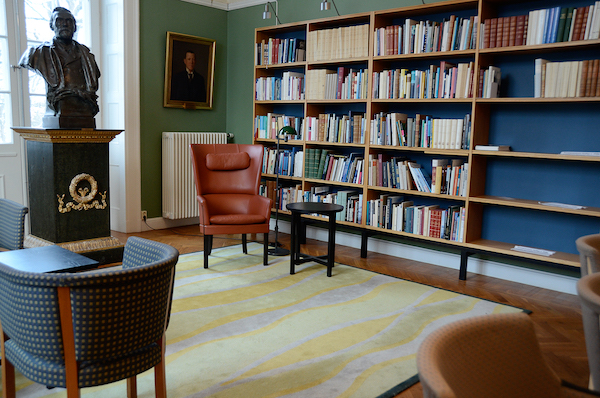}
      & \includegraphics[width=0.30\textwidth]{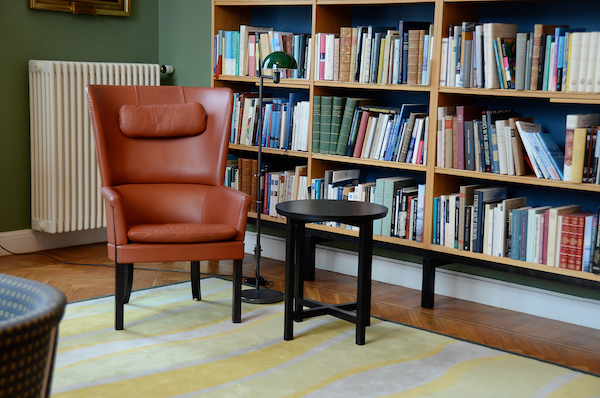}
      & \includegraphics[width=0.30\textwidth]{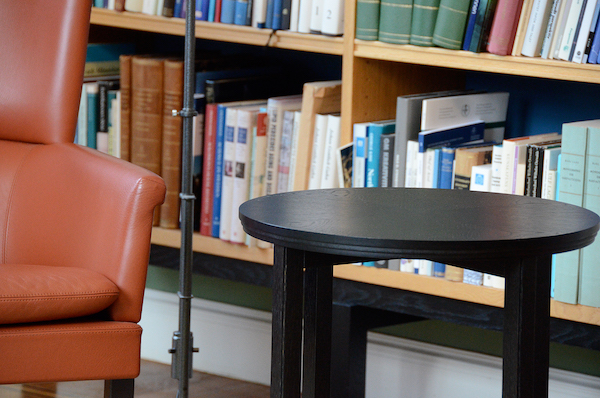} \\    
       \includegraphics[width=0.30\textwidth]{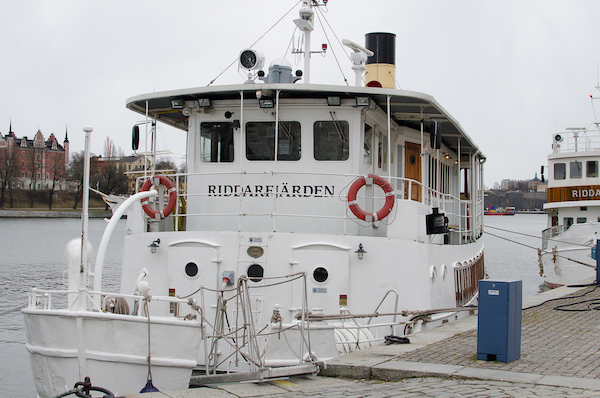}
      & \includegraphics[width=0.30\textwidth]{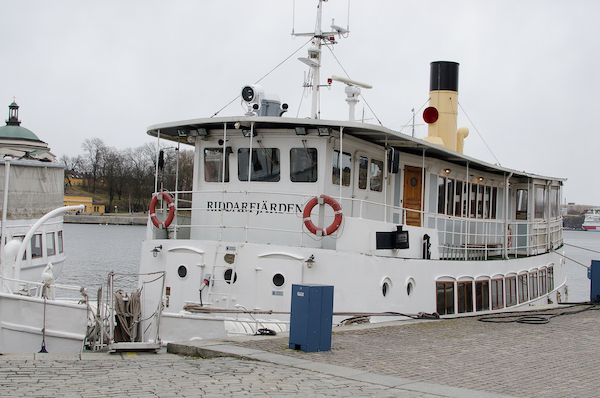}
      & \includegraphics[width=0.30\textwidth]{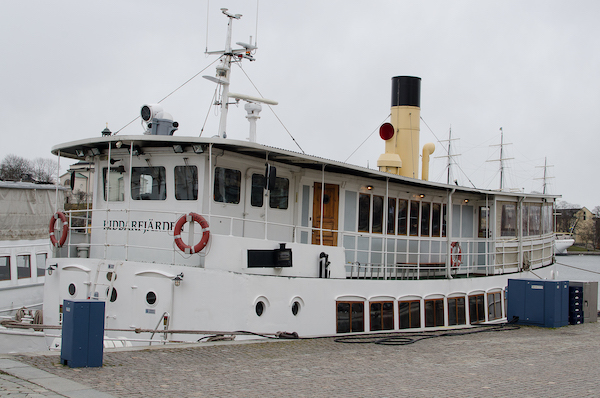} \\
      \includegraphics[width=0.30\textwidth]{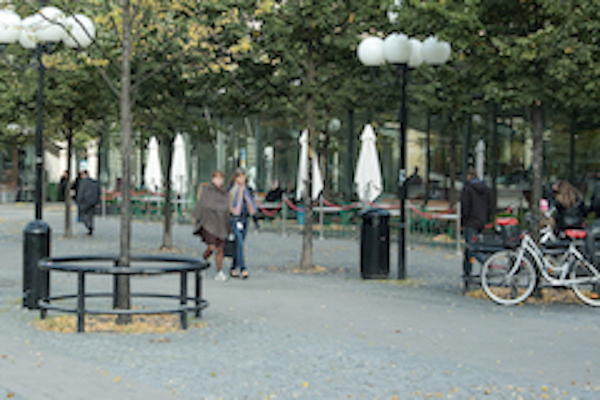}
      & \includegraphics[width=0.30\textwidth]{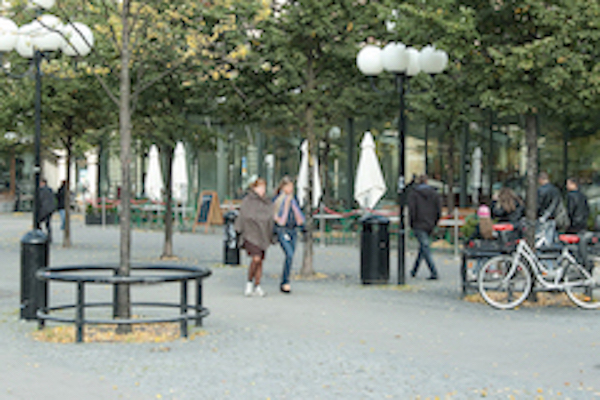}
      & \includegraphics[width=0.30\textwidth]{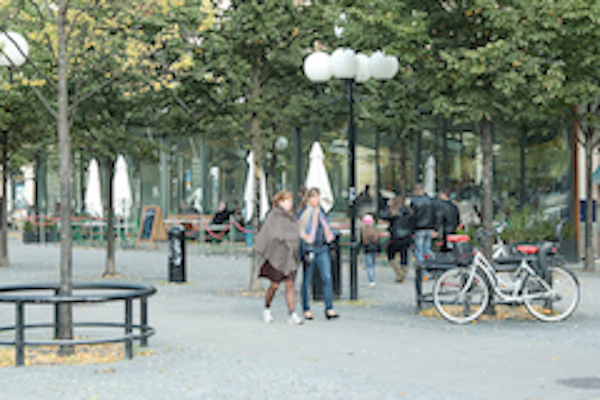}
     \end{tabular}
  \end{center}
  \caption{Due to the effects of natural image transformations, the
    perspective projections of spatial objects and spatio-temporal
    events in the world may appear in substantially different ways
    depending on the viewing conditions. This figure shows images from
    natural scenes under variations in (top row) the viewing distance,
    (middle row) the viewing direction and (bottom row) the relative motion
    between objects in the world and the observer.
    By approximating the non-linear perspective mapping by a local
    linearization (the derivative), these geometric image transformations
    can, to first order of approximation, be modelled by spatial
    scaling transformations, spatial affine transformations and Galilean
   transformations.}
  \label{fig-ill-img-transf-nat-imgs}
\end{figure}

The image and video data, that a vision system is exposed to, is
subject to geometric image transformations, due to variations in the viewing
distance, the viewing direction and the relative motion of objects
in the world relative to the observer. These natural image
transformations do, in turn, cause a substantial
variability, by which an object or event in the world may appear in different
ways to a visual observer:
\begin{itemize}
\item
   Variations in the
   distance between objects in the world and the observer lead to
   variations in scale, often up to over several orders of magnitude,
   which, to first order of approximation of the
   perspective mapping, can be modelled as (uniform) spatial scaling
   transformations (see Figure~\ref{fig-ill-img-transf-nat-imgs} top row).
\item
   Variations in the viewing direction between the
   object and the observer will lead to a wider class of local image
   deformations, with different amount of foreshortening in different spatial
   directions, which, to first order of approximation, can be modelled as
   local spatial affine transformations, where the monocular foreshortening will depend
   upon the slant angle of a surface patch, and correspond to different
   amount of scaling in different directions, also complemented by a
  skewing transformations (see Figure~\ref{fig-ill-img-transf-nat-imgs} middle row).
\item
   Variations in the relative motion between objects in
   the world and the viewing direction will additionally
   transform the joint spatio-temporal video domain in a way that, to first
   order of approximation, can be modelled as local Galilean transformations.
  (see Figure~\ref{fig-ill-img-transf-nat-imgs} bottom row).
\end{itemize}
In this paper, we will study the transformation properties of receptive field%
\footnote{Whereas the notion of a visual receptive field
  traditionally refers to the region in the visual field over which a
  neuron responds to visual stimuli (Hartline
  \citeyear{Har38-AmJPhysio}), we here complement that definition to
  also comprise the computational function of the neuron over the part
  in the visual field in which it reacts to visual patterns.}
responses in the primary visual cortex (V1) under these classes of geometric image
transformations, as well as for temporal scaling transformations
that correspond to objects in the world that move as well as
spatio-temporal events that occur faster or slower.
We will also study the transformation properties of neurons in the
lateral geniculate nucleus (LGN) under a lower variability over spatial scaling
transformations, spatial rotations, temporal scaling transformations and Galilean transformations.
An overall message that we will propose is that if
the family of visual receptive fields is covariant  (or equivariant%
\footnote{In the deep learning literature, the property that we refer
  to as ``covariance'' is often referred to as ``equivariance''. In
  this paper, we use the term ``covariance'', because of the traditional
  use of this terminology in physics, and to maintain consistency with the previous work in
  scale-space theory that this paper builds upon. An operator
  $O$ is said to be covariant under a transformation group $T_p$
  with parameter $p$, if the operator essentially commutes with
  the transformation group, in the sense that
  $O'(T_p(f)) = T_p(O(f)$ for some
  possibly transformed operator $O'$ within the same family of
  operators as $O$.})
under these classes of natural
image transformations, as the family of generalized Gaussian
derivative based receptive fields that we will study is, then these covariance properties make it
possible for the vision system to, up to first order of approximation,
match the receptive field responses under the huge variability caused
by these geometric image transformations, which, in turn, will make it possible for
the vision system to infer more accurate cues to the structure of the
environment.

We will then use these theoretical results to express
predictions concerning biological receptive fields in the primary
visual cortex and the lateral geniculate nucleus, to characterize to
what extent the variability of receptive field
shapes, in terms of receptive fields tuned to different orientations,
motion directions as well as spatial and temporal scales, as has been found by
neurophysiological cell recordings of biological neurons
(DeAngelis {\em et al.\/}\ \citeyear{DeAngOhzFre95-TINS,deAngAnz04-VisNeuroSci};
Ringach \citeyear{Rin01-JNeuroPhys,Rin04-JPhys};
Conway and Livingstone \citeyear{ConLiv06-JNeurSci};
Johnson {\em et al.\/}\ \citeyear{JohHawSha08-JNeuroSci};
De and Horwitz \citeyear{DeHor21-JNPhys}),
could be explained by the primary visual cortex computing a
covariant image representation over the basic classes
of natural geometric image transformations.

\subsection{The importance of covariant receptive fields to handle the
  influence of natural image transformations}
\label{sec-imp-img-transf-rec-field-resp}

For spatial receptive fields, that integrate spatial
    information over non-infinitesimal regions over image space, the
    specific way that the receptive fields accumulate evidence over these
    non-infinitesimal support regions will crucially determine how
    well they are able to handle the huge variability of visual
    stimuli caused by geometric image transformations.

\begin{figure}[hbtp]
  \begin{center}
    \begin{tabular}{cc}
      {\em\small Non-covariant receptive fields\/} & {\em\small Covariant receptive fields\/} \\
      \includegraphics[width=0.40\textwidth]{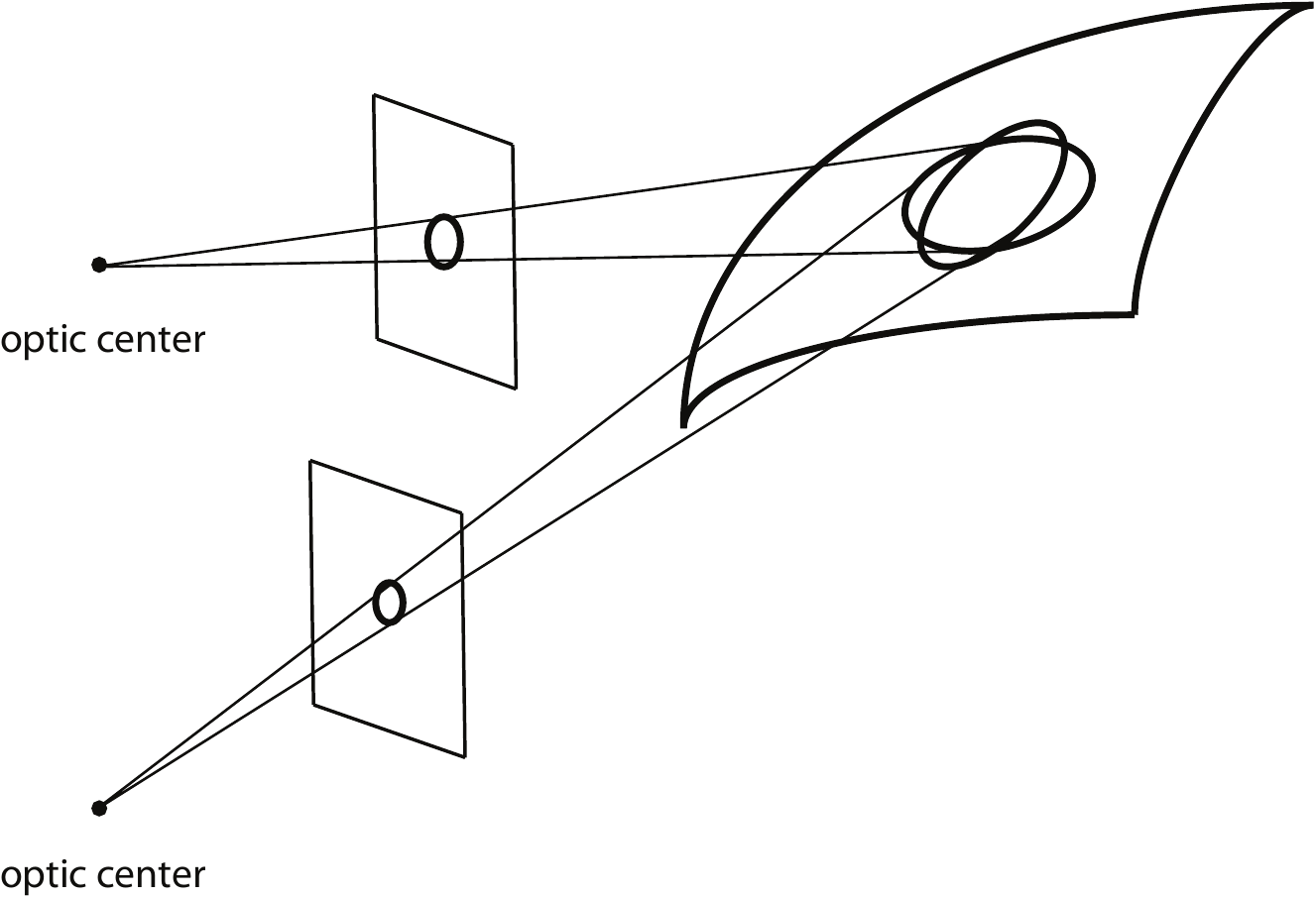}
      & \includegraphics[width=0.40\textwidth]{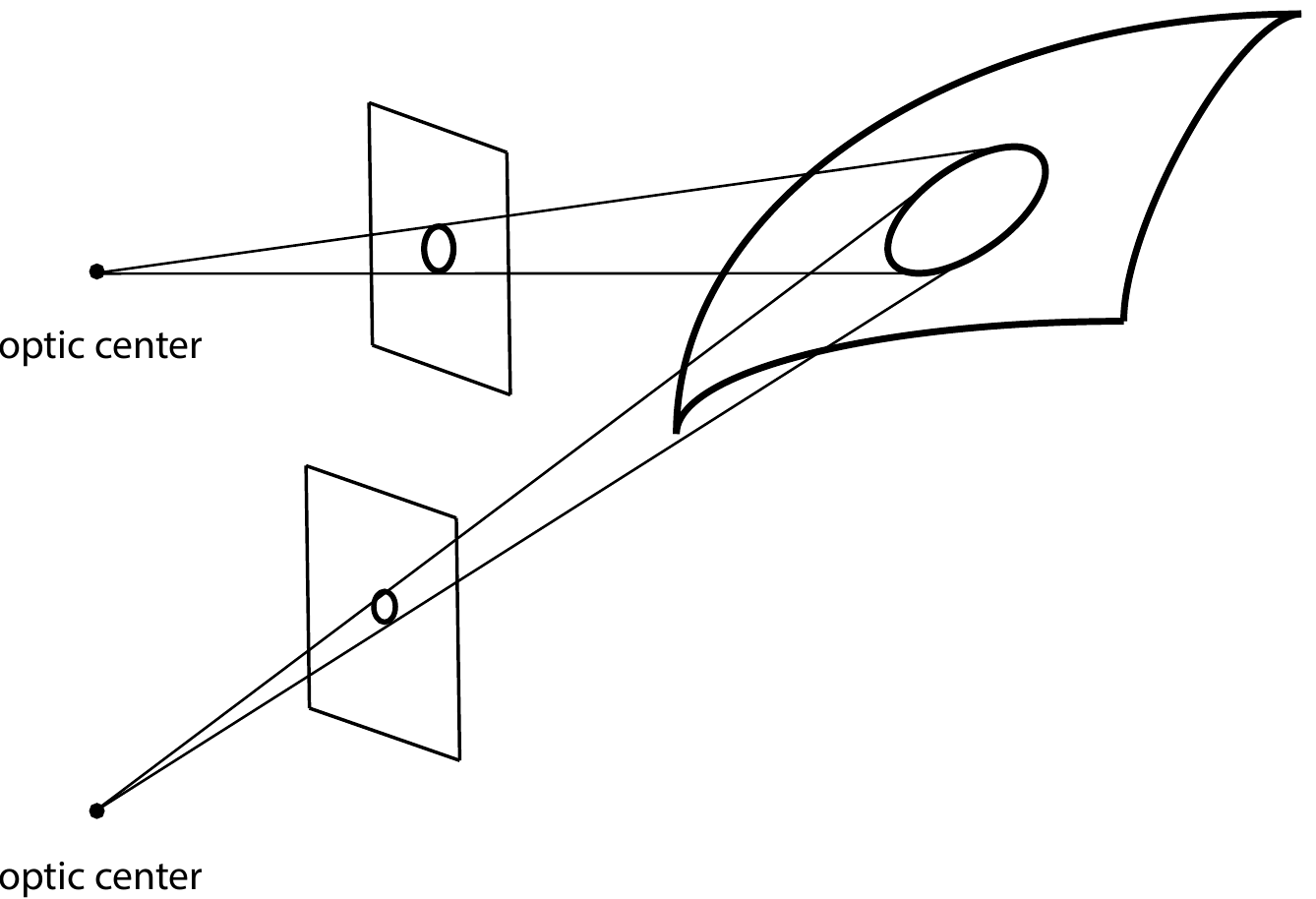} \\
    \end{tabular}
  \end{center}
  \caption{Illustration of backprojected receptive fields for a visual
  observer that observes a smooth surface from two different viewing
  directions in the cases of (left) non-covariant receptive fields and
(right) covariant receptive fields. When using non-covariant receptive
fields, the backprojected receptive fields will not match, which may
cause problems for higher level visual processes that aim at computing
estimates of local surface orientation, whereas when using covariant
receptive fields, the backprojected receptive fields can be adjusted
to match, which in turn enable more accurate estimates of local
surface orientation. (See the text in
Section~\ref{sec-imp-img-transf-rec-field-resp}
for a more detailed explanation.)} 
  \label{fig-backproj-rfs-rot-symm-vs-aff}
\end{figure}

To illustrate the importance of natural image transformation with
respect to image measurements in terms of receptive fields, consider a
visual observer that views a 3-D scene from two different viewing
directions as shown in Figure~\ref{fig-backproj-rfs-rot-symm-vs-aff}.

Let us initially assume that the receptive fields for the two
different viewing conditions are based on image measurements over rotationally
    symmetric support regions in the two image
  domains only. (More precisely, with respect to the theory that is to be
  developed next, concerning receptive fields that are defined from
  spatial derivatives of spatial smoothing kernels, the first crucial
  factor in this model concerns the support regions of the underlying
  spatial smoothing kernels of the spatial receptive fields in the two
  image domains.)

If we backproject these rotationally symmetric support
regions in the two image domains to the tangent plane of the surface in
the world, then these backprojections will, to first order of
approximation, be ellipses. Those ellipses will, however, not coincide
in the tangent plane of the surface, implying that if we try to make
use of the difference between the image measurements over the two
spatial image domains, then there will be an unavoidable source of
error, caused by the difference between the backprojected receptive
fields, which will substantially affect the accuracy when computing cues to
the structure of the world.

If we instead allow ourselves to consider different shapes of the
support regions in the two image domains, and also match the parameters
that determine their shapes such that they coincide in the tangent
plane of the surface, then we can, on the other hand, to first order of
approximation, reduce the mismatch source to the error. For the affine
Gaussian derivative model for spatial receptive fields developed in
this article, that matching property is achieved by matching the
spatial covariance matrices of the affine Gaussian kernels, in such a
way that the support regions of the affine Gaussian kernels may be
ellipses in the respective image domains, and may, for example,
in the most simple case correspond to circles in the tangent plane of
the surface.

Of course, a
complementary problems then concerns how to match the actual values of the
receptive field parameters in a particular viewing situation. That
problem can, however, be seen as a complementary more algorithmic task, in
contrast to the fundamental problem of eliminating an otherwise
inescapable geometric source of
error. In (Lindeberg and G{\aa}rding \citeyear{LG96-IVC}) one example
of such a computational algorithm was developed for the task of
estimating local surface orientation from either monocular or
binocular cues. It was demonstrated that it was possible to design an
iterative procedure for successively adapting the shapes of the
receptive fields, in such a way that it reduced the reconstruction error by about an
order of magnitude after a few (2-3) iterations.

While this specific
example for illustrating the influence of the backprojected support
regions of the receptive fields, when deriving cues to the structure of
the world from local image measurements, is developed
for the case of spatial affine transformations under
a binocular or multi-view viewing situation, a similar geometric
problem regarding the backprojected support regions of the receptive
fields arises also under monocular projection, as well as for
spatio-temporal processing under variations in the relative motion
between objects or events in the world and the observer, then
concerning the backprojections of the spatio-temporal receptive fields
in the 2+1-D video domain to the surfaces of the objects embedded in the 3+1-D world.

A main subject of this paper is to describe a theory for covariant
receptive fields under natural image transformations, which makes it
possible to perfectly match the backprojected receptive field
responses under natural image transformations, approximated by spatial
scaling transformations, spatial affine transformations, Galilean
transformations and temporal scaling transformations.
Since these image transformations may be present in essentially every
natural imaging situation, they constitute essential components to
include in models of visual perception.

\subsection{Theory of covariant visual receptive fields}

The receptive fields
of the neurons in the visual pathway constitute the main computational primitives in the early stages of the visual sensory system. These visual receptive fields integrate and process visual information over local regions over space and time, which is then passed on to higher layers. Capturing the functional properties of visual receptive fields, as well as trying to explain what determines their computational function, constitute key ingredients in understanding vision.

In this work, we will for our theoretical studies build upon
the Gaussian derivative model for
visual receptive fields, which was initially theoretically proposed by
Koenderink and van Doorn (\citeyear{Koe84,KoeDoo87-BC,KoeDoo92-PAMI}),
used for modelling biological receptive fields by
Young and his co-workers
(\citeyear{You87-SV,YouLesMey01-SV,YouLes01-SV})
and then generalized by Lindeberg (\citeyear{Lin13-BICY,Lin21-Heliyon}).
With regard to the topic of the tematic collection on ``Symmetry as a
Guiding Principle in Artificial and Brain Neural Networks'', the
subject of this article is to describe how symmetry properties in
terms of covariance
properties under natural image transformations constitute
an essential component in that normative theory of visual receptive
fields, as well as how such covariance properties may be important
with regard to biological vision, specifically to understand the organization of the
receptive fields in the primary visual cortex.

 
It will be shown that a purely spatial version of the
studied generalized Gaussian derivative model for visual receptive
fields allows for spatial scale covariance and spatial affine
covariance, implying that it can, to first order of approximation,
handle variations in the distance between objects in the world and the
observer, as well as first-order approximations of the geometric transformations
induced by viewing the surface of a smooth 3-D object from
different viewing distances and viewing directions in the world. It
will also be shown that for a more general spatio-temporal version
of the resulting generalized Gaussian derivative model, the covariance
properties do in addition extend to local Galilean transformations, which
makes it possible to perfectly handle first-order approximations
of the geometric transformations induced by viewing objects in the
world under different relative motions between the object and the
observer. The spatio-temporal receptive fields in this model do
also obey temporal scale covariance, which makes it possible
to handle objects that move as well as
spatio-temporal events that occur faster or slower
in the world. In these ways, the resulting model for visual receptive
fields respects the main classes of geometric image
transformations in vision.

We argue that if the goal is to build realistic computational models
of biological receptive fields, or more generally a computational
model of a visual system
that should be able to handle general classes of natural image or video data
in a robust and stable manner, it is
essential that the internal visual representations obey sufficient covariance
properties under these classes of geometric image transformations, to, in turn, make it possible to
achieve geometric invariance properties at the systems level of the visual system.

We conclude by using predictions from the presented theory to describe
implications for biological vision. Specifically, we state
experimentally testable hypotheses to explore to what extent the
variabilities of receptive field shapes in the primary visual cortex
span corresponding variabilities as described by the basic classes of
natural image transformations, in terms of spatial scaling
transformations, spatial affine transformations, Galilean
transformations and temporal scaling transformations. These hypotheses
are, in turn, intended for experimentalists to explore and
characterize to what extent the primary
visual cortex computes a covariant representation of receptive field
responses over those classes of natural image transformations.

In this context, we do also describe theoretically how receptive field responses at
coarser levels of spatial and temporal scales can be computed from
finer scales using cascade smoothing properties over spatial and
temporal scales, meaning that it could be sufficient for the visual
receptive fields in the first layers to only implement the receptive
fields at the finest level of spatial and temporal scales, from which
coarser scale representations could then be inferred at higher levels in
the visual hierarchy.

\section{Methods}

\subsection{The generalized Gaussian derivative model for spatial and
  spatio-temporal receptive fields}

In this section, we will describe the generalized Gaussian derivative model
for linear receptive fields, in the cases of either (i)~image data defined
over a purely spatial domain, or (ii)~video data defined over a joint
spatio-temporal domain, including brief conceptual overviews of how
this model can be derived in a principled axiomatic manner, from symmetry
properties in relation to the first layers of the visual hierarchy.
We do also give pointers to previously established results that
demonstrate how these models do qualitatively very well model biological
receptive fields in the lateral geniculate nucleus (LGN) and the
primary visual cortex (V1), as established by comparisons to results of
neurophysiological recordings of visual neurons. This theoretical
background will then constitute the theoretical background for the
material in Section~\ref{sec-results}, concerning covariance
properties of the visual receptive fields according to the Gaussian
derivative model, and the implications of such covariance properties for modelling
and explaining the families of receptive field shapes found in biological vision.

\subsubsection{Purely spatial models for linear receptive fields}
\label{sec-spat-gauss-der-model}

For image data defined over a 2-D spatial domain with image coordinates
$x = (x_1, x_2)^T \in \bbbr^2$,
an axiomatic derivation in (Lindeberg \citeyear{Lin10-JMIV}; Theorem~5), based on the
assumptions of linearity, translational
covariance, semi-group%
\footnote{A semi-group structure over convolutions with a family of
  kernels $T(x_1, x_2;\; s)$, where $s$ is the parameter of the
  semi-group, means that the result of convolving two kernels with
  each other will be a kernel of the same family under addition of the
  parameters, {\em i.e.\/},
  $T(\cdot, \cdot;\; s_1) * T(\cdot, \cdot;\;
s_2) = T(\cdot, \cdot;\; s_1+s_2)$. Note that a convolution structure
is implied from the previous assumptions of linearity and translation covariance.}
structure over scale and non-creation of new structures from finer to coarser
levels of scales in terms of non-enhancement of local extrema%
\footnote{Non-enhancement of local extrema means that the value of the
  image representation at any local maximum over the image domain must not
  increase from finer to coarser levels of scales, and that the value
  of the image representation at any local minimum over the image domain
  must not decrease.},
combined with certain regularity assumptions, shows that under
evolution of a spatial scale parameter $s$ that reflects the spatial
size of the receptive fields, the spatially smoothed
image representations $L(x_1, x_2;\; s)$ that underlie the output from the receptive
fields must satisfy a spatial diffusion equation of the form
\begin{equation}
  \label{eq-vel-adapt-scsp-diff-eq-spat}
  \partial_s L 
  = \frac{1}{2} \,
      \nabla_{(x_1,x_2)}^T \left( \Sigma_{(x_1, x_2)} \, \nabla_{(x_1,x_2)} L \right)
      - \delta_{(x_1, x_2)}^T \nabla_{(x_1,x_2)} L 
\end{equation}
with initial condition $L(x_1, x_2;\; 0) = f(x_1, x_2)$, where $f(x_1, x_2)$ is
the input image, $\nabla_{(x_1,x_2)} = (\partial_{x_1}, \partial_{x_2})^T$ is the
spatial gradient vector, $\Sigma_{(x_1, x_2)}$ is a spatial covariance
matrix and $\delta_{(x_1,x_2)}$ is a spatial drift vector.

The physical interpretation of this equation is that if we think of
the intensity distribution $L$ in the image plane as a heat distribution,
then this equation describes how the heat distribution will evolve over
the virtual time variable $s$, where hot spots will get successively 
cooler and cold spots will get successively warmer. The evolution
will in this sense serve as a spatial smoothing process over the
variable $s$, which we henceforth will refer to as a spatial scale
parameter of the receptive fields.

The spatial covariance matrix $\Sigma_{(x_1, x_2)}$ in this equation determines
how much spatial smoothing is performed in different directions in the
spatial domain, whereas the spatial drift vector $\delta_{(x_1,x_2)}$
implies that the smoothed image structures may move spatially over the
image domain during the smoothing process. The latter property can,
for example, be used for aligning image structures under variations in
the disparity field for a binocular visual observer. For the rest of
the present treatment, we will, however, focus on monocular viewing
conditions and disregard that term in the case of purely spatial image
data. For the case of spatio-temporal data, to be considered later, a
correspondence to this drift term will on the other hand be used for
handling motion, to perform spatial smoothing of objects that
change the positions of their projections in the image plane
over time, without causing excessive amounts of motion blur, if
desirable, as it will be for a receptive field tuned to a particular
motion direction.

\begin{figure}[hbtp]
  \begin{center}
    \begin{tabular}{cccccc}
      \hspace{-4mm}
     \includegraphics[width=0.16\textwidth]{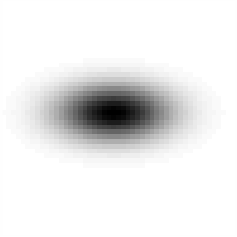} \hspace{-4mm} &
      \includegraphics[width=0.16\textwidth]{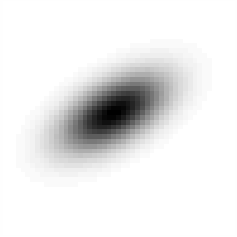} \hspace{-4mm} &
      \includegraphics[width=0.16\textwidth]{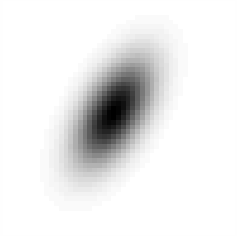} \hspace{-4mm} &
      \includegraphics[width=0.16\textwidth]{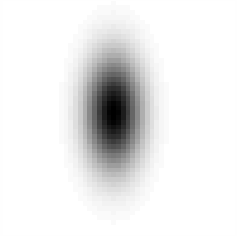} \hspace{-4mm} &
      \includegraphics[width=0.16\textwidth]{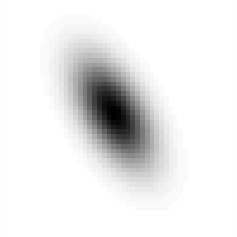} \hspace{-4mm} &
      \includegraphics[width=0.16\textwidth]{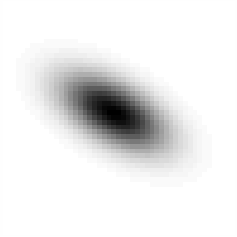} \hspace{-4mm} \\
     \includegraphics[width=0.16\textwidth]{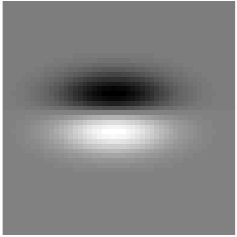} \hspace{-4mm} &
      \includegraphics[width=0.16\textwidth]{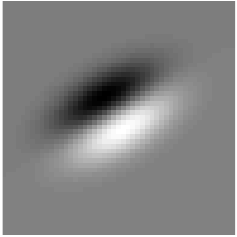} \hspace{-4mm} &
      \includegraphics[width=0.16\textwidth]{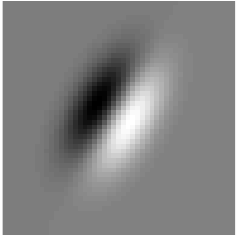} \hspace{-4mm} &
      \includegraphics[width=0.16\textwidth]{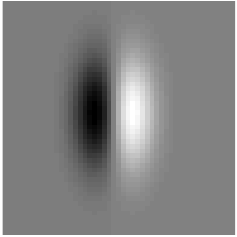} \hspace{-4mm} &
      \includegraphics[width=0.16\textwidth]{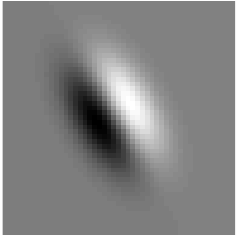} \hspace{-4mm} &
      \includegraphics[width=0.16\textwidth]{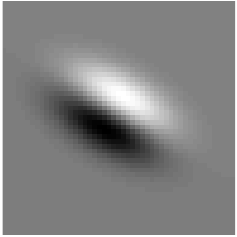} \hspace{-4mm} \\
    \includegraphics[width=0.16\textwidth]{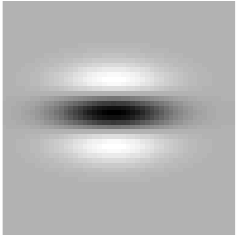} \hspace{-4mm} &
      \includegraphics[width=0.16\textwidth]{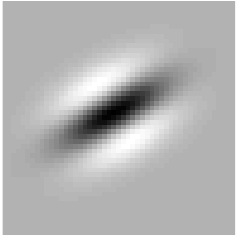} \hspace{-4mm} &
      \includegraphics[width=0.16\textwidth]{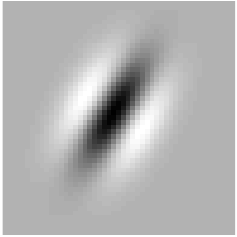} \hspace{-4mm} &
      \includegraphics[width=0.16\textwidth]{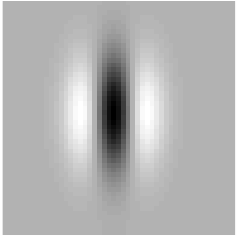} \hspace{-4mm} &
      \includegraphics[width=0.16\textwidth]{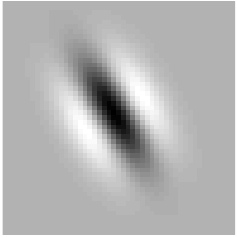} \hspace{-4mm} &
      \includegraphics[width=0.16\textwidth]{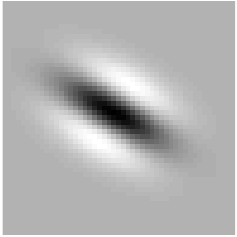} \hspace{-4mm} \\
     \hspace{-4mm}
    \end{tabular} 
  \end{center}
  \vspace{-4mm}
  \caption{Illustration of the variability of receptive field shapes
    over spatial rotations and the order of spatial differentiation.
    This figure shows affine Gaussian kernels $g(x_1, x_2;\; \Sigma)$ with their directional
    derivatives $\partial_{\varphi} g(x_1, x_2;\; \Sigma)$ and
    $\partial_{\varphi\varphi}g(x_1, x_2;\; \Sigma)$ 
    up to order two,
    here with the eigenvalues of $\Sigma$ being
    $\lambda_1 = 64$, $\lambda_2=16$ and for the image orientations
    $\varphi = 0, \pi/6, \pi/3, \pi/2, 2\pi/3, 5\pi/6$.
    With regard to the classes of image transformations considered in
    this paper, this figure shows an expansion over the rotation
    group. In addition, the family of affine Gaussian kernels also
    comprises an expansion over a variability over the ratio between
    the eigenvalues of the spatial covariance matrix $\Sigma$, that
    determines the shape of the affine Gaussian kernels, see
    Figure~\ref{fig-aff-Gauss-hemisphere} for complementary illustrations.
   (Horizontal dimension: $x \in [-24, 24]$. Vertical dimension: $y \in [-24, 24]$.)}
  \label{fig-aff-elong-filters-dir-ders}
\end{figure}

In terms of the spatial smoothing
kernels that underlie the definition of a family of spatial receptive
fields, that integrate information over a spatial support region in the
image domain, this smoothing process corresponds to smoothing
with spatially shifted affine Gaussian kernels of the form
\begin{equation}
  \label{eq-gauss-gen-spattemp}
  g(x;\; \Sigma_s, \delta_s) 
   = \frac{1}{2 \pi \sqrt{\det \Sigma_s}} \,
      e^{- {(x - \delta_s)^T \Sigma_s^{-1} (x - \delta_s)}/{2}},
    \end{equation}
for a scale dependent spatial covariance matrix $\Sigma_s$ and a
scale-dependent spatial drift vector $\delta_s$ of the form
$\Sigma_s = s \, \Sigma_{(x_1, x_2)}$ and $\delta_s = s \, \delta_{(x_1,x_2)}$.
Requiring these spatial smoothing kernels to additionally be mirror symmetric around the
origin, and removing the parameter $s$ from the notation, does, in turn,
lead to the regular family of affine Gaussian kernels of the form
\begin{equation}
  \label{eq-aff-gauss}
  g(x;\; \Sigma) 
   = \frac{1}{2 \pi \sqrt{\det \Sigma}} \,
      e^{-x^T \Sigma^{-1} x/2}.
\end{equation}
Considering that spatial derivatives of the output of affine Gaussian convolution
also satisfy the diffusion equation (\ref{eq-vel-adapt-scsp-diff-eq-spat}),
and incorporating scale-normalized derivatives of the form
(Lindeberg \citeyear{Lin97-IJCV}) 
\begin{equation}
  \partial_{\xi^{\alpha}}
   = s^{(\alpha_1 + \alpha_2)/m} \,
       \partial_{x_1}^{\alpha_1} \, \partial_{x_2}^{\alpha_2},
 \end{equation}
where $\alpha = (\alpha_1, \alpha_2)$ denotes the order of differentiation,
as well as combining multiple partial derivatives into directional
derivative operators according to 
\begin{equation}
  \partial_{\varphi}^m
  = (\cos \alpha \, \partial_{x_1} + \sin \alpha \, \partial_{x_2})^m,
\end{equation}
where we choose the directions $\varphi$ for computing the directional
derivative to be aligned with the directions of the eigenvectors of the
spatial covariance matrix $\Sigma$, we obtain the following canonical model%
\footnote{In the model below, the parameters $s$ and $\Sigma$ are not
  independent, implying that we could multiply the spatial scale
  parameter $s$ with some constant $C$, while dividing the spatial
  covariance matrix $\Sigma$ with the same constant, and still get the
same spatial receptive field. To handle this problem, we can
introduce a convention to normalize $\Sigma$ in some way, for
example, by choosing a normalization such that the maximum eigenvalue
of $\Sigma$ is equal to 1. For a monocular view of smooth surface
under orthographic projection, such a normalization is closely related to
the foreshortening transformation from the tangent plane of the
surface to the image plane. If the slant angle of the surface is
$\sigma$ relative to the viewing direction, then a rotationally
symmetric receptive field, when backprojected to the tangent plane of
the surface, would then correspond to the minimum eigenvalue of
$\Sigma$ being $\cos^2 \sigma$.}
for oriented spatial receptive fields of the form
(Lindeberg \citeyear{Lin21-Heliyon}; Equation~(23))
\begin{equation}
  \label{eq-spat-RF-model}
  T_{simple}(x_1, x_2;\; s, \varphi, \Sigma, m)
  = T_{{\varphi}^{m},norm}(x_1, x_2;\; s, \Sigma)  
  = s^{m/2} \, \partial_{\varphi}^{m} \left( g(x_1, x_2;\; s \Sigma) \right).
\end{equation}
Figure~\ref{fig-aff-elong-filters-dir-ders} shows examples of such
affine Gaussian kernels with their directional derivatives
up to order two, for different orientations
in the image domain, expressed for one specific choice of
the ratio between the eigenvalues of the spatial covariance matrix
$\Sigma$. More generally, this model also comprises variations of the
ratio between the eigenvalues, as illustrated in Figure~\ref{fig-aff-Gauss-hemisphere}(left),
which visualizes a uniform distribution of zero-order affine Gaussian
kernels on a hemisphere. The receptive fields in the latter
illustration
should then, in turn, be complemented by
directional derivatives of such kernels up to a given order of spatial
differentiation, see Figure~\ref{fig-aff-Gauss-hemisphere}(right) for
an illustration of such directional derivative receptive fields of
order one.
 In the most idealized version of the theory, one could think of
   all these receptive fields, with their directional derivatives up to
   a certain order, as being present at every point in the
   visual field (if we disregard the linear increase in minimal receptive
   field size from the fovea towards the periphery in a foveal vision system).
With respect to a specific implementation of this model in a specific
vision system, it then constitutes a complementary design choice, to
sample the variability of that parameter space in an efficient manner.

For rotationally symmetric receptive fields over the spatial domain,
a corresponding model can instead be expressed of the form
(Lindeberg \citeyear{Lin21-Heliyon}; Equation~(39))
\begin{equation}
  \label{eq-lgn-model-spat}
  T_{LGN}(x_1, x_2\; s) 
  = \pm s \, (\partial_{x_1 x_1} + \partial_{x_2 x_2}) \, g(x_1, x_2;\; s).
\end{equation}
In (Lindeberg \citeyear{Lin13-BICY}) it is proposed that the purely spatial
component of simple cells in
the primary visual cortex can be modelled by directional derivatives
of affine Gaussian kernels of the form (\ref{eq-spat-RF-model});
see Figures~16 and~17 in
(Lindeberg \citeyear{Lin21-Heliyon}) for illustrations. It is also
proposed that the purely spatial component of LGN neurons can be modelled by
Laplacians of Gaussians of the form (\ref{eq-lgn-model-spat}); see Figure~13 in
(Lindeberg \citeyear{Lin21-Heliyon}).

The affine Gaussian derivative model of simple cells in
(\ref{eq-spat-RF-model}) goes
beyond the previous biological modelling work by Young (\citeyear{You87-SV})
and Koenderink and van Doorn's (\citeyear{KoeDoo87-BC,KoeDoo92-PAMI})
theoretical studies,
in that the underlying spatial smoothing kernels in our model are anisotropic, as
opposed to isotropic in Young's and Koenderink and van Doorn's models.
This anisotropy leads to better
approximation of the biological receptive fields, which are more
elongated (anisotropic) over the spatial image domain than can
be accurately captured based on derivatives of
rotationally symmetric Gaussian kernels. The generalized Gaussian
derivative model based on affine Gaussian kernels does also enable
affine covariance, as opposed to mere scale and rotational covariance
for the regular Gaussian derivative model, based on derivatives of the
rotationally symmetric Gaussian kernel.

\subsubsection{Joint spatio-temporal models for linear receptive fields}

For video data defined over a 2+1-D non-causal spatio-temporal domain
with coordinates $p = (x_1, x_2, t)^T \in \bbbr^3$,
a corresponding axiomatic derivation, also based on Theorem~5 in
(Lindeberg \citeyear{Lin10-JMIV}), while now formulated over a joint
spatio-temporal domain, shows that under evolution over a
joint spatio-temporal scale parameter $u$, the spatio-temporally smoothed
video representations $L(x_1, x_2, t;\; s, \tau, v, \Sigma)$, that
underlie the output from the receptive
fields, must satisfy a spatio-temporal diffusion equation of the form
\begin{equation}
  \label{eq-vel-adapt-scsp-diff-eq-spat-temp}
  \partial_u L 
  = \frac{1}{2} \,
      \nabla_{(x_1, x_2, t)}^T \left( \Sigma_{(x_1, x_2, t)} \, \nabla_{(x_1, x_2, t)} L \right) -
      \delta_{(x_1, x_2, t)}^T \nabla_{(x_1, x_2, t)} L 
\end{equation}
for $u$ being some convex combination of the spatial scale parameter
$s$ and the temporal scale parameter $\tau$, and with initial condition
$L(x_1, x_2, t;\; 0, 0, v, \Sigma) = f(x_1, x_2, t)$, where $f(x_1, x_2, t)$ is
the input video, $\nabla_{(x_1, x_2, t)} = (\partial_{x_1}, \partial_{x_2}, \partial_t)^T$
is the spatio-temporal gradient, $\Sigma_{(x_1,x_2,t)}$ is a spatio-temporal covariance
matrix and $\delta_{(x_1,x_2,t)}$ is a spatio-temporal drift vector.

If we think of the intensity distribution $L$ over
joint space-time as
a heat distribution, then this equation describes how the heat distribution
will evolve over an additional virtual time variable $u$ (operating at a higher
meta level than the physical time variable $t$), with the spatio-temporal
covariance matrix $\Sigma_{(x_1,x_2,t)}$ describing how much smoothing
of the virtual heat distribution will be performed in different
directions in joint (real) space-time, whereas the spatio-temporal drift vector
$\delta_{(x_1,x_2,t)}$ describes how the image structures may move in
joint (real) space-time as function of the virtual variable $u$, which
is important for handling the perspective projections of objects that move in the real world.

In (Lindeberg \citeyear{Lin21-Heliyon}; Appendix~B.1) it is shown that the
solution of this equation can be expressed as the convolution with
spatio-temporal kernels of the form
\begin{equation}
  \label{eq-spat-temp-RF-model}
  T(x_1, x_2, t;\; s, \tau, v, \Sigma) 
  = g(x_1 - v_1 t, x_2 - v_2 t;\; s \, \Sigma) \, h(t;\; \tau),
\end{equation}
where $\Sigma$ is a spatial covariance matrix,
$v = (v_1, v_2)$ is a velocity vector and $h(t;\; \tau)$ is a 1-D
temporal Gaussian kernel. Combining this model with spatial
directional derivatives in a similar way as for the spatial model, 
and introducing scale-normalized temporal derivatives (Lindeberg
\citeyear{Lin16-JMIV}) of the form
\begin{equation}
  \partial_{\zeta}^n = \tau^{n/2} \, \partial_t^n
\end{equation}
as well as corresponding velocity-adapted temporal derivatives
according to
\begin{equation}
  \partial_{\bar t} = v_1 \, \partial_{x_1} + v_2 \, \partial_{x_2} + \partial_t,
\end{equation}
then leads to a non-causal
spatio-temporal receptive field model of the form
(Lindeberg \citeyear{Lin21-Heliyon}; Equation~(32))
\begin{align}
  \begin{split}
  \label{eq-spat-temp-RF-model-der-norm-non-caus}
    & T_{{\varphi}^m, {\bar t}^n, norm}(x_1, x_2, t;\; s, \tau, v, \Sigma)
     = s_{\varphi}^{m/2} \, 
          \tau^{n/2} \, 
         \partial_{\varphi}^{m} \, \partial_{\bar t}^n 
          \left( g(x_1 - v_1 t, x_2 - v_2 t;\; s \, \Sigma) \, h(t;\; \tau) \right).
  \end{split}
\end{align}
Regarding the more realistic case of video data defined over a 2+1-D time-causal
spatio-temporal domain, in which the future cannot be accessed,
theoretical arguments in (Lindeberg \citeyear{Lin16-JMIV}) and
(Lindeberg \citeyear{Lin21-Heliyon}; Appendix~B.2; see Equation~(33)) lead to
spatio-temporal smoothing kernels of a similar form
as in (\ref{eq-spat-temp-RF-model-der-norm-non-caus}),
\begin{align}
  \begin{split}
    \label{eq-spat-temp-RF-model-der-norm-caus}
    T_{simple}(x_1, x_2, t;\; s, \tau, \varphi, v, \Sigma, m, n) =
     T_{{\varphi}^m, {\bar t}^n, norm}(x_1, x_2, t;\; s, \tau, v, \Sigma)
  \end{split}\nonumber\\
  \begin{split}
   &  = s_{\varphi}^{m/2} \, 
          \tau^{n/2} \, 
         \partial_{\varphi}^{m} \,\partial_{\bar t}^n 
          \left( g(x_1 - v_1 t, x_2 - v_2 t;\; s \, \Sigma) \,
            \psi(t;\; \tau, c) \right),
  \end{split}
\end{align}
but now
with the non-causal temporal Gaussian kernel replaced by a time-causal temporal kernel
referred to as the time-causal limit kernel
\begin{equation}
  \label{eq-time-caus-limit-kern}
  h(t;\; \tau) = \psi(t;\; \tau, c),
\end{equation}
defined by having a Fourier transform of the form
\begin{equation}
  \label{eq-FT-comp-kern-log-distr-limit}
     \hat{\Psi}(\omega;\; \tau, c) 
     = \prod_{k=1}^{\infty} \frac{1}{1 + i \, c^{-k} \sqrt{c^2-1} \sqrt{\tau} \, \omega},
\end{equation}
and corresponding the an infinite set of truncated exponential kernels
coupled in cascade, with specifically chosen time constants to
obtain temporal scale covariance (Lindeberg
\citeyear{Lin16-JMIV} Section~5; \citeyear{Lin23-BICY} Section~3.1).

Figures~\ref{fig-non-caus-sep-spat-temp-rec-fields}
and~\ref{fig-non-caus-vel-adapt-spat-temp-rec-fields} show examples of
such spatio-temporal receptive fields up to order two, for the case of a 1+1-D
spatio-temporal domain.
In Figure~\ref{fig-non-caus-sep-spat-temp-rec-fields}, where the image
velocity $v$ is zero, the receptive fields are space-time separable,
whereas in Figure~\ref{fig-non-caus-vel-adapt-spat-temp-rec-fields},
where the image velocity is non-zero, the receptive fields are not
separable over space-time. In the most idealized version of the theory, we
could think of these velocity-adapted receptive fields, for all
velocity values within some range, as being present at
every point in the image domain, see
Figure~\ref{fig-distr-vel-adapt-spattemp-kernels-1+1-D} for an
illustration.

In this
way, the receptive field family will be Galilean covariant,
which makes it possible for the vision
system to handle observations of moving objects and events in the world,
irrespective of the relative motion between the object or the event in the world and
the viewing direction.

For rotationally symmetric receptive fields over the spatial domain,
a corresponding model can instead be expressed of the form
(Lindeberg \citeyear{Lin21-Heliyon}; Equation~(39)) 
\begin{equation}
  \label{eq-lgn-model-spat-temp}
  T_{LGN}(x_1, x_2, t;\; s, \tau) 
  = \pm s \, \tau^{n/2} \,
  (\partial_{x_1 x_1} + \partial_{x_2 x_2}) \, g(x_1, x_2;\; s) \,
  \partial_{t^n} \psi(t;\; \tau, c).
\end{equation}
In (Lindeberg \citeyear{Lin16-JMIV}) it is proposed that simple cells in
the primary visual cortex can be modelled by spatio-temporal kernels
of the form (\ref{eq-spat-temp-RF-model-der-norm-caus})
for $h(t;\; \tau) = \psi(t;\; t)$ according to
(\ref{eq-FT-comp-kern-log-distr-limit}); see Figure~18  in
(Lindeberg \citeyear{Lin21-Heliyon}) for illustrations. It is also
proposed that lagged and non-lagged
LGN neurons can be modelled by temporal derivatives of
Laplacians of Gaussians of the form (\ref{eq-lgn-model-spat-temp}); see Figure~12 in
(Lindeberg \citeyear{Lin21-Heliyon}).

These models go beyond the previous modelling work by Young and his
co-workers (\citeyear{YouLesMey01-SV,YouLes01-SV}) in that
our models are truly time-causal, as opposed to
Young's non-causal model, and also that the parameterization of the
filter shapes in our model is different, more in line with the geometry of the
spatio-temporal image transformations.

\subsubsection{Spatio-chromatic and spatio-chrom-temporal receptive field
  models}

Both of the above spatial and spatio-temporal models can be extended
from operating on pure grey-level image information to operating on
colour image data, by instead applying these models to each channel of a
colour-opponent colour space; see (Lindeberg \citeyear{Lin13-BICY,Lin21-Heliyon})
for further details as well as modelling results. Since the geometric
image transformations have the same effect on colour-opponent channels
as on grey-level information, we do here in this article, without loss of generality, develop
the theory for the case of grey-level images only, while noting that
similar algebraic transformation properties will hold for all the
channels in a colour-opponent representation of the image or video data.

\section{Results}
\label{sec-results}

\subsection{Covariance properties under geometric image transformations}

In this section, we will describe covariance properties of the above
described generalized Gaussian derivative model, in the cases of
either (i)~spatial receptive fields over a purely spatial domain, or
(ii)~spatio-temporal receptive fields over a joint spatio-temporal
domain. 

An initial treatment of this topic, in less developed form, was given in the supplementary
material of (Lindeberg \citeyear{Lin21-Heliyon}), however, on
a format that may not be easy to digest for a reader with
neuroscientific background.
In the present treatment, we develop the
covariance properties of the spatial and spatio-temporal receptive
field models in an extended as well as more explicit manner,
that should be more easy to access. Specifically, we incorporate the transformation
properties of the full derivative-based receptive field models as opposed to
only the transformation properties of the spatial or spatio-temporal
smoothing processes in the previous treatment,
and also comprising more developed interpretations of these covariance properties
with regard to the associated variabilities in image and video data
under natural image transformations.
Then, in Section~\ref{sec-impl-biol-vision} we will use these
theoretical results to formulate predictions with regard to
implications for biological vision, including the formulation of a
set of testable biological hypotheses as well as needs for
characterizing the distributions of biological receptive field shapes.

It will be shown that the purely spatial model for simple cells in
(\ref{eq-spat-RF-model}) is covariant under {\em affine image
transformations\/} of the form
\begin{equation}
  \label{eq-aff-transf}
  x_R = \left( \begin{array}{c} x_{R,1} \\ x_{R,2} \end{array} \right) 
  = \left( \begin{array}{cc} a_{11} & a_{12} \\ a_{21} &  a_{22} \end{array} \right)
      \left( \begin{array}{c} x_{L,1} \\ x_{L,2} \end{array} \right)
  = A \, x_L,
\end{equation}
which includes the special cases of covariance under
{\em spatial scaling transformations\/} with
\begin{equation}
  \label{eq-sc-transf}  
   A = S_x \, I = \left( \begin{array}{cc} S_x & 0 \\ 0 &  S_x \end{array} \right)
\end{equation}
and {\em spatial rotations\/} with
\begin{equation}
  A = R =
  \left( \begin{array}{cc}
           \cos \theta & -\sin \theta \\
           \sin \theta &  \cos \theta
         \end{array}
   \right).
 \end{equation}
 For the spatial model of LGN neurons (\ref{eq-lgn-model-spat}),
 affine covariance cannot be
 achieved, because the underlying spatial smoothing kernels are
 rotationally symmetric. That model does instead obey covariance under
 spatial scaling transformations and spatial rotations.

 It will also be shown that the joint spatio-temporal model of
 simple cells in (\ref{eq-spat-temp-RF-model-der-norm-caus})
 is, in addition to affine covariant, also covariant under
 {\em Galilean transformations\/} of the form
 \begin{equation}
   \label{eq-gal-transf}
   p' = \left( \begin{array}{c} x'_1 \\ x'_2 \\ t' \end{array} \right)
   = \left(
          \begin{array}{ccc}
            1 & 0 & u_1 \\
            0 & 1 & u_2 \\
            0 & 0 & 1
          \end{array}
        \right)
        \left( \begin{array}{c} x_1 \\ x_2 \\ t \end{array} \right)
        = \left(
          \begin{array}{c}
            x_1 + u_1 t \\
            x_2 + u_2 t \\
            t
          \end{array}
          \right)
        = G \, p
\end{equation}
as well as covariant under {\em temporal scaling transformations\/}
\begin{equation}
  \label{eq-temp-sc-transf}
  t' = S_t \, t.
\end{equation}
For the joint spatio-temporal model of LGN neurons
(\ref{eq-lgn-model-spat-temp}), the covariance properties are,
however, restricted to spatial scaling transformations, spatial
rotations and temporal scaling transformations.%
\footnote{It is, however, possible to complement the spatio-temporal
  LGN model (\ref{eq-lgn-model-spat-temp}) by velocity-adaptation,
  leading to a receptive field model of the form $h_{LGN}(x_1, x_2, t;\; s, \tau, v) 
  = \pm s \, \tau^{n/2} \,
  (\partial_{x_1 x_1} + \partial_{x_2 x_2}) \, \partial_{{\bar t}^n}
  (g(x-v_1 t, y-v_2 t;\; s) \, \, \psi(t;\; \tau, c))$, to also obtain
  Galilean covariance. Notably, there are neurophysiological results indicating
that there also exist neurons in the LGN that are sensitive to motion
directions (Ghodrati {\em et al.\/}
\citeyear{GhoKhaLeh17-ProNeurobiol} Section~1.4.2).}

The real-world significance of these covariance properties is that:
\begin{itemize}
\item
  If we linearize the perspective mapping from a local surface patch
  on a smooth surface in the world to the image plane, then the
  deformation of a pattern on the surface to the image plane can, to
  first order of approximation,
  be modelled as a local affine transformation of the form (\ref{eq-aff-transf}).
\item
  If we linearize the projective mapping between two views of a
  surface patch from different viewing directions, then the projections of
  a pattern on the surface to the two images can, to first order of
  approximation, be related by an affine
  transformation of the form (\ref{eq-aff-transf}).
\item
  If we view an object or a spatio-temporal event in the world in two
  situations, that correspond to different relative velocities in
  relation to the viewing direction, then the two spatio-temporal
  image patterns can, to first order of approximation, be
  related by local Galilean transformations of the
  form~(\ref{eq-gal-transf}).
\item
  If an object moves or an event in the world occurs faster or slower, then the
  spatio-temporal image patterns arising from that moving object
  or event are related by a
  temporal scaling transformation of the form (\ref{eq-temp-sc-transf}).
\end{itemize}
By the family of receptive fields being covariant under these image
transformation implies that the vision system will have the potential
ability to perfectly match the output from the receptive field
families, given two or more views of the same object or event in the
world. In this way, the vision system will have the potential of
substantially reducing the measurement errors, when inferring cues
about properties in the world from image measurements in terms of
receptive fields. For example, in an early work on affine covariant
and affine invariant receptive field representations based on affine
Gaussian kernels, it was demonstrated that it is possible
to design algorithms that reduce the error in estimating local surface
orientation from monocular or binocular cues by an order of magnitude,
compared to using receptive fields based on the rotationally symmetric
Gaussian kernel (Lindeberg and G{\aa}rding \citeyear{LG96-IVC}; see
the experimental results in Tables~1--3).

Based on this theoretical reasoning, in combination with previous
experimental support, we argue that it is essential for an artificial
or biological vision system to obey sufficient covariance properties,
or sufficient approximations thereof, in order to robustly handle the
variabilities in image and video data caused by the natural image
transformations that arise when viewing objects and events in a complex
natural environment.

\subsubsection{Formalism to be used}

When applying an image transformation between two image domains, not
only the spatial or the spatio-temporal representations of the
receptive field output need to be transformed,
but also the parameters for the receptive field models
need to be appropriately matched. In other words, it is necessary to
also derive the transformation properties of the
spatial scale parameter $s$ and the spatial covariance matrix $\Sigma$
for the spatial model, and additionally the transformation properties
of the temporal scale parameter
$\tau$ and the image velocity $v$ for the spatio-temporal model.

In the following subsections, we will derive the covariance properties of the
receptive field models under the respective classes of image transformations.
To highlight how the parameters are also affected by the image
transformations, we will therefore
express these covariance properties in terms of explicit image
transformations regarding the output from the receptive field models,
complemented by explicit commutative diagrams.
To reduce the complexity of the expressions, we will focus on the
transformation properties of the output from the underlying spatial and
spatio-temporal smoothing kernels only.

The output from the actual derivative-based
receptive fields can then be obtained by also complementing the
transformation properties of the output from the spatial and spatio-temporal smoothing
operations with the
transformation properties of the respective spatial and temporal derivative
operators, where spatial derivatives transform according to
\begin{equation}
  (\nabla_{(x_{L,1}, x_{L,2})} L_L)(x_L) =A^T (\nabla_{(x_{R,1}, x_{R,2})} L_R)(x_R)
\end{equation}
under an affine transformation (\ref{eq-aff-transf}) of the form 
\begin{equation}
   L_L(x_L) = L_R(x_R)  \quad\quad \mbox{for} \quad\quad x_R = A \, x_L,
\end{equation}
whereas spatio-temporal derivatives transform according to
\begin{equation}
  (\nabla_{(x_1, y_1,t)} L)(p) =G^T (\nabla_{(x_1', x_2',t')} L')(p')
\end{equation}
under a Galilean transformation (\ref{eq-gal-transf}) of the form 
\begin{equation}
   L(p) = L'(p')  \quad\quad \mbox{for} \quad\quad p' = G \, p
 \end{equation}
with $p = (x_1, x_2, t)^T$ and $p' = (x'_1, x'_2, t')^T$.

By necessity, the treatment that follows next in this section will be somewhat
technical. The hasty reader may skip the mathematical derivations in
the following text and instead focus on the resulting
commutative diagrams in
Figures~\ref{fig-comm-diag-aff-transf-spat}--\ref{fig-comm-diag-sc-transf-spat-temp},
which describe the
essence of the transformation properties of the receptive fields
in the generalized Gaussian derivative model.

\subsubsection{Covariance properties for the purely spatial receptive fields}

\paragraph{Transformation property under spatial affine
  transformations}
\label{sec-aff-cov-spat}

To model the essential effect of the receptive field output in terms
of the spatial smoothing transformations applied to two spatial image patterns
$f_L(x_L)$ and $f_R(x_R)$, that are related according to a spatial affine
transformation (\ref{eq-aff-transf}) of the form
\begin{equation}
  x_R = A \, x_L,
\end{equation}
let us define the corresponding spatially smoothed representations
$L_L(x_L;\; s, \Sigma_L)$ and $L_R(x_L;\; s, \Sigma_R)$ obtained by
convolving $f_L(x_L)$ and $f_R(x_R)$ with affine Gaussian kernels of the form
(\ref{eq-aff-gauss}), using spatial scale parameters $s_L$ and $s_R$ as well as
spatial covariance matrices $\Sigma_L$ and $\Sigma_R$, respectively.

Then, according to (Lindeberg and G{\aa}rding \citeyear{LG96-IVC};
Section~4.1), these spatially smoothed representations%
\footnote{In the area of scale-space theory, these spatially smoothed
  image representations are referred to as (here affine) spatial scale-space
  representations.}
are related
according to
\begin{equation}
  L_L(x_L;\; s, \Sigma_L) = L_R(x_R;\; s, \Sigma_R)
\end{equation} 
for
\begin{equation}
  \Sigma_R = A \Sigma_L A^T,
\end{equation} 
which constitutes an affine covariant transformation property, as illustrated in the
commutative diagram in Figure~\ref{fig-comm-diag-aff-transf-spat}.
Specifically, the affine
covariance property implies that for every
receptive field output from the image $f_L(x_L)$ computed with spatial
covariance matrix $\Sigma_L$, there exists a corresponding spatial
covariance matrix $\Sigma_R$, such that the receptive field output from
the image $f_R(x_R)$ using that covariance matrix can be perfectly
matched to the receptive field output from image $f_L(x_L)$ using the spatial
covariance matrix $\Sigma_L$.

This property thus means
that the family of spatial receptive fields will, up to first order of approximation, have the
ability to handle the image deformations induced between the perspective
projections of multiple views of a smooth local surface patch in the world.

For the spatial LGN model in (\ref{eq-lgn-model-spat}), a weaker
rotational covariance property holds, implying that under a spatial
rotation
\begin{equation}
  x_R = R \, x_L,
\end{equation}
where $R$ is a 2-D rotation matrix,
the corresponding spatially smoothed representations are related
according to
\begin{equation}
  L_L(x_L;\; s) = L_R(x_R;\; s).
\end{equation} 
This property implies that image structures arising from projections
of objects that rotate around an axis aligned with the viewing
direction are handled in a structurally similar manner.

\paragraph{Transformation property under spatial scaling transformations}
\label{sec-sc-cov-spat}

In the special case when the spatial affine transformation represented
by the
affine matrix $A$ reduces to a scaling transformation, having scaling
matrix $S$ with scaling factor $S_x$ according to (\ref{eq-sc-transf} )
\begin{equation}
  x_R = S_x \, x_L,
\end{equation}
the above affine covariance relation reduces to the form
\begin{equation}
  L_R(x_R;\; s_R, \Sigma) = L_L(x_L;\; s_L, \Sigma),
\end{equation}
provided that the spatial scale parameters are related according to
\begin{equation}
   s_R = S_x^2 \, s_L
 \end{equation}
 and assuming that the spatial covariance matrices are the same
\begin{equation}
   \Sigma_R = \Sigma_L = \Sigma.
\end{equation}
With regard to the previous treatment in Section~\ref{sec-spat-gauss-der-model},
this reflects the scale covariance property of the spatial
smoothing transformation underlying the simple cell model in
(\ref{eq-spat-RF-model}), as illustrated in the commutative diagram in
Figure~\ref{fig-comm-diag-spat-scaling}.

A corresponding spatial scale covariance property does also hold for
the spatial smoothing component in the
spatial LGN model in (\ref{eq-lgn-model-spat}), and is of the form
\begin{equation}
  L_R(x_R;\; s_R) = L_L(x_L;\; s_L),
\end{equation}
provided that the spatial scale parameters are related according to
\begin{equation}
   s_R = S_x^2 \, s_L.
 \end{equation}
This scale covariance property
implies that the spatial receptive field family will, up to first order of
approximation, have the ability to handle the image deformations
induced by viewing an object from different distances relative to the
observer, as well as handling objects in the world that have similar
spatial appearance, while being of different physical size.

\subsubsection{Covariance properties for the joint spatio-temporal receptive
  fields}

\paragraph{Transformation property under Galilean transformations}
\label{sec-gal-cov}

To model the essential effect of the receptive field output in terms
of the spatio-temporal smoothing transformation applied to two
spatio-temporal video patterns
$f(p)$ and $f'(p')$, that are related according to a Galilean
transformation (\ref{eq-gal-transf}) according to
\begin{equation}
  f'(x', t') = f(x, t)
\end{equation}
for $t' = t'$ and
\begin{equation}
  \label{eq-gal-trans-sec-cov-prop}
    x' = x + u t,
\end{equation}
let us define the corresponding spatio-temporally smoothed video
representations%
\footnote{In the area of scale-space theory, these spatio-temporally
  smoothed video representations are referred to as spatio-temporal
  scale-space representations.}
$L(x, t;\; s, \tau, v, \Sigma)$ and $L'(x', t';\; s', \tau', v', \Sigma')$ obtained by
convolving $f(p)$ and $f'(p)$ with spatio-temporal smoothing kernels of the form
(\ref{eq-spat-temp-RF-model}) using spatial scale parameters $s$ and $s'$,
temporal scale parameters $\tau$ and $\tau'$,
velocity vectors $v$ and $v'$, and
spatial covariance matrices $\Sigma$ and $\Sigma'$, respectively:
\begin{align}
  \begin{split}
    \label{eq-spattemp-kern-def-cov-prop-1}
    L(x, t;\; s, \tau, v, \Sigma) & = T(x, t;\; s, \tau, v, \Sigma) * f(x, t),
  \end{split}\\
  \begin{split}
    \label{eq-spattemp-kern-def-cov-prop-2}
    L'(x', t';\; s', \tau', v', \Sigma') & = T(x', t';\; s', \tau', v', \Sigma') * f(x', t').
  \end{split}
\end{align}
By relating these two representations to each other, according to a
change of variables, it then follows that the
spatio-temporally smoothed video representations are related
according to the Galilean covariance property
\begin{equation}
  L'(x', t';\; s', \tau', v', \Sigma') = L(x, t;\; s, \tau, v, \Sigma),
\end{equation}
provided that the velocity parameters for the two different receptive
field models are related according to
\begin{equation}
  v' = v + u
\end{equation}
and assuming that the other receptive field parameters are the same,
{\em i.e.\/}, that $s' = s$, $\Sigma' = \Sigma$ and $\tau' = \tau$,
see the commutative diagram in Figure~\ref{fig-comm-diag-gal-transf}
for an illustration.

The regular LGN model (\ref{eq-lgn-model-spat}),
based on space-time separable receptive fields,
is not covariant under Galilean transformations. Motivated by
the fact that there are neurophysiological results showing that
that there also exist neurons in the LGN that are sensitive to motion
directions (Ghodrati {\em et al.\/}
\citeyear{GhoKhaLeh17-ProNeurobiol} Section~1.4.2), we could,
however, also formulate a tentative Galilean covariant receptive field
model for LGN neurons,
having rotationally symmetric receptive fields over the spatial
domain, according to
\begin{equation}
  \label{eq-lgn-model-spat-temp-vel}
  h_{LGN}(x_1, x_2, t;\; s, \tau, v) 
  = \pm s \, \tau^{n/2} \,
  (\partial_{x_1 x_1} + \partial_{x_2 x_2}) \, \partial_{{\bar t}^n}
  (g(x-v_1 t, y-v_2 t;\; s) \, \, \psi(t;\; \tau, c)).
\end{equation}
Then, the underlying purely spatio-temporally smoothed representations,
disregarding the Laplacian operator
$\nabla_{x_1,x_2)}^2 =\partial_{x_1 x_1} + \partial_{x_2 x_2}$
and the temporal derivative operator $\partial_{{\bar t}^n}$, will 
under a Galilean transformation (\ref{eq-gal-trans-sec-cov-prop})
be related according to
\begin{equation}
  L'(x', t';\; s', \tau', v') = L(x, t;\; s, \tau, v),
\end{equation}
provided that the velocity parameters for the two different receptive
field models are related according to
\begin{equation}
  v' = v + u,
\end{equation}
and assuming that the other receptive field parameters are the same,
{\em i.e.\/}, that $s' = s$ and $\tau' = \tau$.

These two Galilean covariance properties mean that the corresponding
families of spatio-temporal receptive
fields will, up to first order of approximation, have the ability to
handle the image deformations induced between objects that move, as
well as
spatio-temporal events that occur, with different relative velocities
between the object/event and the observer.

\paragraph{Transformation property under temporal scaling transformations}

To model the essential effect of the receptive field output in terms
of the spatio-temporal smoothing transformations applied to two
spatio-temporal video patterns
$f(p)$ and $f'(p')$, that are related according to a temporal scaling
transformation (\ref{eq-temp-sc-transf}) according to
\begin{equation}
  f'(x', t') = f(x, t)
\end{equation}
for $x' = x$ and
\begin{equation}
   t' = S_t^2 \, t,
\end{equation}
with the temporal scaling factor $S_t$ restricted%
\footnote{The reason why the temporal scaling factor is here restricted to
  being an integer power of the distribution parameter of the
  time-causal limit kernel, is that the temporal scale levels resulting from
  convolution with this temporal smoothing kernel are genuinely discrete. Thereby, exact
  temporal scale covariance can only be obtained for temporal scaling
  factors that perfectly match the available ratios between the temporal scale
  levels in the corresponding temporal multi-scale representation. For
  other scaling factors, the results will instead only be numerical
  approximations, whose accuracy depends on the density of the discrete
  temporal scale levels, as determined by the distribution parameter
  $c >1$ of the time-causal limit kernel.}
to being an integer%
\footnote{If the time-causal limit kernel in the general form
  (\ref{eq-spat-temp-RF-model}) of spatio-temporal
  smoothing kernels, that this treatment is based on, is replaced by a non-causal
  temporal Gaussian kernel, then a corresponding temporal scale
  covariance will also hold, however then instead over a continuum of temporal scaling
  factors $S_t$. as opposed to a discrete subset as for the time-causal model.}
power of the distribution parameter $c > 1$ of
the time-causal limit kernel $\psi(t;\; \tau, c)$ in the time-causal
spatio-temporal receptive field model
\begin{equation}
   S_t = c^i,
\end{equation}
let us analogously to above define the corresponding spatio-temporally
smoothed video representations 
$L(x, t;\; s, \tau, v, \Sigma)$ and $L'(x', t';\; s', \tau', v', \Sigma', v')$
of $f(p)$ and $f'(p')$, respectively,
according to (\ref{eq-spattemp-kern-def-cov-prop-1})
and (\ref{eq-spattemp-kern-def-cov-prop-2}).

Then, due to the scale-covariant property of the time-causal limit
kernel (Lindeberg \citeyear{Lin16-JMIV} Equation~(47);
\citeyear{Lin23-BICY} Equation~(34)),
these spatio-temporally smoothed video representations are related
according to the temporal scale covariance property
\begin{equation}
  L'(x', t';\; s', \tau', v', \Sigma') = L(x, t;\; s, \tau, v, \Sigma),
\end{equation}
provided that the temporal scale parameters and the velocity parameters
for the two receptive field models are matched according to
\begin{equation}
  \tau' = S_t^2 \, \tau  \quad\quad\mbox{and}\quad\quad v' = v/S,
\end{equation}
and assuming that the other receptive field parameters are the same,
{\em i.e.\/}, that $s' = s$ and $\Sigma' = \Sigma$,
see the commutative diagram in Figure~\ref{fig-comm-diag-temp-sc-transf}
for an illustration.

A corresponding temporal covariance property does also hold for the
space-time separable spatio-temporal LGN model in
(\ref{eq-lgn-model-spat-temp}), of the form
\begin{equation}
  L'(x, t';\; s, \tau')   = L_L(x, t;\; s, \tau),
\end{equation}
provided that the temporal scale parameters are related according to
\begin{equation}
   \tau' = S_t^2 \, \tau.
 \end{equation}

These temporal scaling covariance properties mean
that the families of spatio-temporal receptive fields
will be able to handle objects that move and events that
occur faster or slower in the world.

\paragraph{Transformation property under spatial affine
  transformations}

To model the essential effect of the receptive field output in terms
of the spatio-temporal smoothing transformation applied to two
spatio-temporal video patterns
$f(p_L)$ and $f(p_R)$, that are related according to a spatial affine
transformation (\ref{eq-aff-transf}) according to
\begin{equation}
  x_R = A \, x_L,
\end{equation}
let us analogously to in Section~\ref{sec-gal-cov} define the
corresponding spatio-temporally
smoothed video representations
$L_L(x_L, t_L;\; s_L, \tau_L, v_L, \Sigma_L)$
and $L_R(x_R, t_R;\; s_R, \tau_T, v_R, \Sigma_R)$
of $f(p_L) = f(p)$ and $f(p_R) = f'(p)$, respectively, according to
(\ref{eq-spattemp-kern-def-cov-prop-1}) and
(\ref{eq-spattemp-kern-def-cov-prop-2}).

Then, based on similar transformation properties as are used for
deriving the affine covariance property of the purely spatial receptive
field model in Section~\ref{sec-aff-cov-spat}, it follows that
these spatio-temporally smoothed video representations are related
according to the spatial affine covariant transformation property
\begin{equation}
  L_R(x_R, t_R;\; s_R, \tau_R, v_R, \Sigma_R) = L_L(x_L, t_L;\; s_L, \tau_L, v_L, \Sigma_L) 
\end{equation}
for
\begin{equation}
  \Sigma_R = A \, \Sigma_L \, A^T \quad\mbox{and}\quad v_R = A \, v_L,
\end{equation}
provided that the other receptive field parameters are the same,
{\em i.e.\/}, that $s_R = s_L$, $\tau_R = \tau_L$ and $v_R = v_L$.

With regard to the previous treatment in Section~\ref{sec-spat-gauss-der-model},
this property reflects the spatial affine covariance property of the spatio-temporal 
smoothing transformation underlying the simple cell model in
(\ref{eq-spat-temp-RF-model-der-norm-caus}), 
as illustrated in the commutative diagram in
Figure~\ref{fig-comm-diag-aff-transf-spat-temp}.

This affine covariance property implies that, also under non-zero relative motion
between an object or event in the world and the observer, the family
of spatio-temporal receptive fields will have the ability to, up to
first order of approximation, handle the image deformations caused by
viewing the surface pattern of a smooth surface in the world under
variations in the viewing direction relative to the observer.

For the spatio-temporal LGN model with velocity adaptation
in (\ref{eq-lgn-model-spat-temp-vel}), a weaker
rotational covariance property holds, implying that under a spatial
rotation
\begin{equation}
  x_R = R \, x_L,
\end{equation}
where $R$ is a 2-D rotation matrix,
the corresponding spatially smoothed representations are related
according to
\begin{equation}
  L_L(x_L, t;\; s, \tau, v_L) = L_R(x_R, t;\; s, \tau, v_R),
\end{equation}
provided that
\begin{equation}
  v_R = R \, v_L.
\end{equation}
For the spatio-temporal LGN model without velocity adaptation
in (\ref{eq-lgn-model-spat-temp}), a similar result holds, with the
conceptual difference that $v_R = v_L = 0$.

As for the previously studied purely spatial case, these rotational
covariance properties imply that image structures arising from projections
of objects that rotate around an axis aligned with the viewing
direction are handled in a structurally similar manner.

\paragraph{Transformation property under spatial scaling transformations}

In the special case when the spatial affine transformation represented the
affine matrix $A$ reduces to a scaling transformation to a spatial scaling
matrix $S$, with spatial scaling factor $S_x$ according to (\ref{eq-sc-transf})
\begin{equation}
   x_R = S_x \, x_L,
\end{equation}
the above spatial affine covariance relation reduces to the form
\begin{equation}
  L_R(x_R, t;\; s_R, \tau, v_R, \Sigma)   = L_L(x_L, t;\; s_L, \tau, v_L, \Sigma),
\end{equation}
provided that the spatial scale parameters and the velocity vectors
are related according to
\begin{equation}
   x_R = S_x^2 \, x_L \quad\quad\mbox{and}\quad\quad v_R = S_x \, v,
 \end{equation}
and assuming that the other receptive field parameters are the same,
{\em i.e.\/}, that $\tau_R = \tau_L = \tau$,
$\Sigma_R = \Sigma_L =\Sigma$ and $v_R = v_L$.

With regard to the previous treatment in Section~\ref{sec-spat-gauss-der-model},
this instead reflects the scale covariance property of the spatio-temporal 
smoothing transformation underlying the simple cell model in
(\ref{eq-spat-temp-RF-model-der-norm-caus}), 
as illustrated in the commutative diagram in
Figure~\ref{fig-comm-diag-sc-transf-spat-temp}.

A corresponding spatial scale covariance property does also hold for the
space-time separable spatio-temporal LGN model in
(\ref{eq-lgn-model-spat-temp}), of the form
\begin{equation}
  L_R(x_R, t;\; s_R, \tau)   = L_L(x_L, t;\; s_L, \tau),
\end{equation}
provided that the spatial scale parameters are related according to
\begin{equation}
   x_R = S_x^2 \, x_L.
 \end{equation}
 
Similar to the previous spatial scale covariance property for purely
spatial receptive fields in Section~\ref{sec-sc-cov-spat}, this spatial
covariance property means that the spatio-temporal receptive field
family will, up to first order of
approximation, have the ability to handle the image deformations
induced by viewing an object from different distances relative to the
observer, as well as handling objects in the world that have similar
spatial appearance, while being of different physical size.

\subsection{Implications of the theory for biological vision}
\label{sec-impl-biol-vision}

Concerning biological implications of the presented theory, it is
sometimes argued that the oriented simple cells in the primary visual
cortex serve as mere edge detectors. In view of the presented theory, the
oriented receptive fields of the simple cells can, on the other hand, also be viewed as
populations of
receptive fields, that together make it possible to capture local image
deformations in the image domain, to, in turn, serve as a cue for
deriving cues to surface orientation and surface shape in the world.
In addition, the spatio-temporal dependencies of the simple cells are also
essential to handle objects that move as well as spatio-temporal
events that occur, with possibly different relative motions in relation to the observer.

According to the presented theory, the spatial and spatio-temporal
receptive fields are expanded over their associated filter parameters,
whereby the population of receptive fields becomes able to handle the
different classes of locally linearized image and video deformations.
This is done in such a way that the output from the receptive fields can be
perfectly matched under these image and video transformations, provided that the
receptive field parameters are properly matched to the actual
transformations that occur in a particular imaging situation.
Specifically, an interesting follow-up question of this work to biological vision
research concerns how well biological vision spans corresponding families of
receptive fields, as predicted by the presented theory.

In the study of orientation maps of receptive field families around pinwheels
(Bonhoeffer and Grinvald (\citeyear{BonGri91-Nature}), it
has been found that for higher mammals, oriented receptive fields in the primary
visual cortex are laid out with a specific organization regarding their
directional distribution, in that neurons with preference for a
similar orientation over the spatial domain are grouped together, and in
such a way that the preferred orientation varies continuously around
the singularities in the orientation maps known as pinwheels
(see Figure~\ref{fig-orient-maps}).
First of all, such an organization shows that
biological vision performs an explicit expansion over the group of
spatial rotations, which is a subgroup of the affine group.

Beyond variations in mere orientation selectivity of neurons,
neurophysiological investigations by Blasdel (\citeyear{Bla92-JNeuroSci})
have, however, also shown that the degree of orientation selectivity
varies regularly over the cortex, and is different near
{\em vs.\/} further away from the center of a pinwheel;
see also Nauhaus {\em et al.\/} (\citeyear{NauBenCarRin09-Neuron})
and Koch {\em et al.\/} (\citeyear{KocJinAloZai16-NatComm}).
Specifically, the orientation selectivity is lowest near the positions
of the centers of the pinwheels, and then increases with the distance from
the pinwheel. In view of the spatial model for
simple cells (\ref{eq-spat-RF-model}), such a behaviour would be a characteristic
property, if the spatial receptive fields would be laid out over the
cortex according to a distribution over the spatial covariance matrices
of the affine Gaussian kernels, which determine the purely spatial
smoothing component in the spatial receptive field model.

For the closest to isotropic affine Gaussian kernels, a small
perturbation of the spatial covariance matrix of the
spatial smoothing kernel could cause a larger
shift in the preferred orientation than for a highly anisotropic
Gaussian kernel. The singularity (the pinwheel) in such a model would
therefore correspond to the limit case of a rotationally symmetric
Gaussian kernel, alternatively an affine Gaussian kernel as near as
possible to a rotationally symmetric Gaussian kernel, within some
complementary constraint not modelled here, if the
parameter variation in the biological system does not reach the same
limits as in our idealized theoretical model.

\begin{figure}[hbtp]
   \begin{center}
     \begin{tabular}{cc}
       {\em\small Zero-order affine Gaussian kernels\/}
       & {\em\small First-order affine Gaussian derivative kernels\/} \\
       
       \includegraphics[width=0.42\textwidth]{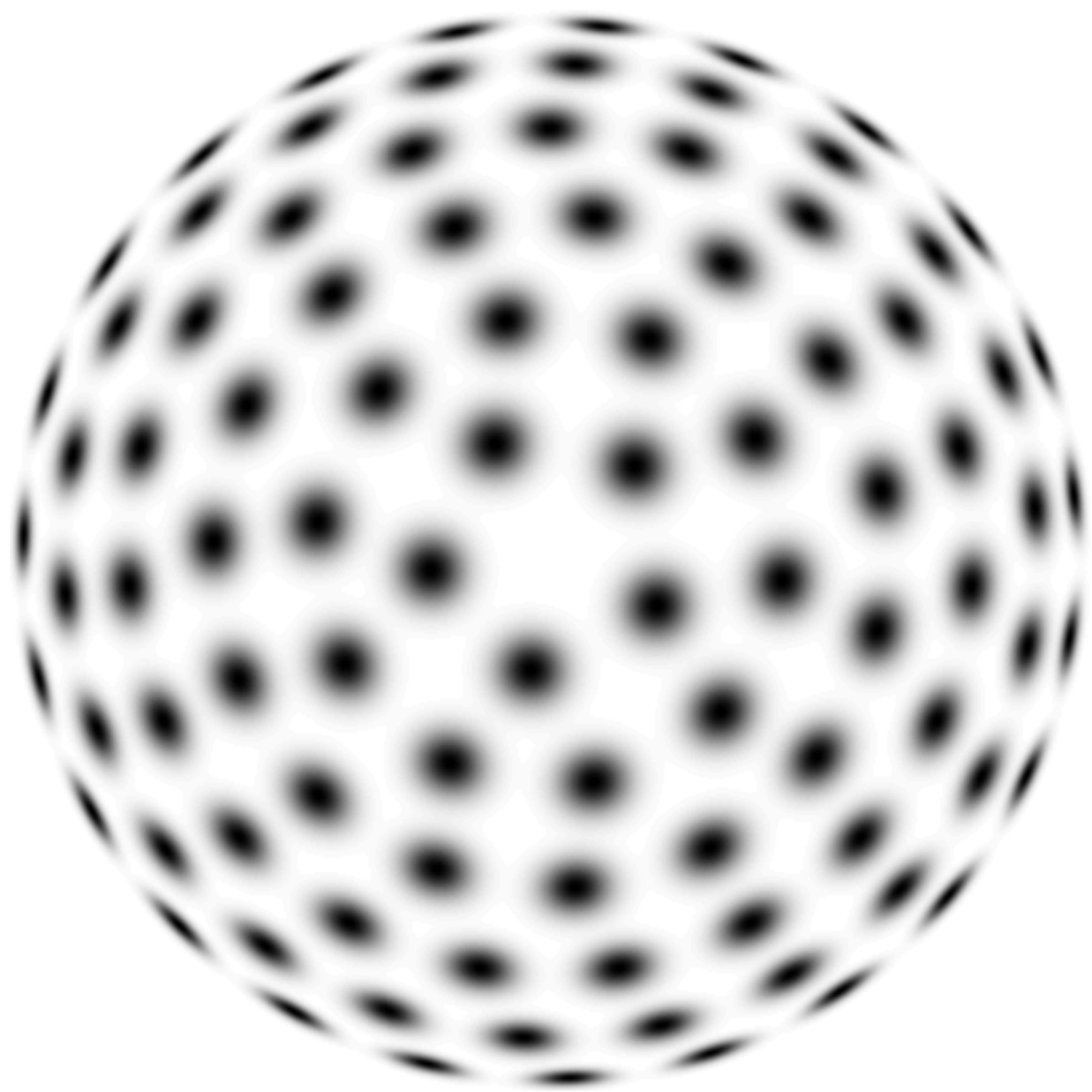} 
       & \includegraphics[width=0.42\textwidth]{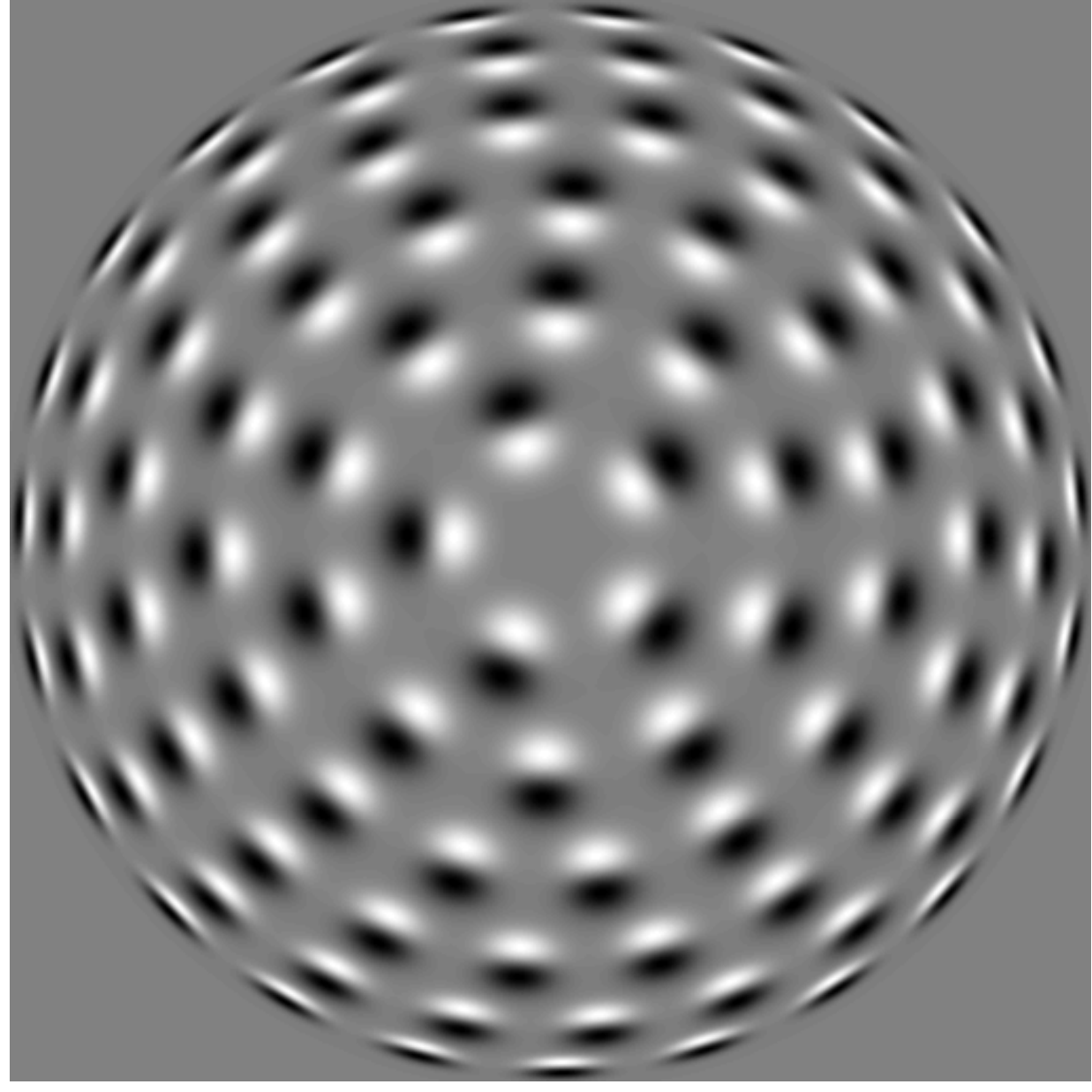} 
     \end{tabular}
   \end{center}
   \caption{Illustration of the variability of receptive field shapes
     over spatial affine transformations that also involve a
     variability of the eccentricity of the receptive fields.
     (left) Zero-order affine Gaussian kernels for different
     orientations $\varphi$ in the image domain as well as for different ratios
     between the eigenvalues $\lambda_1$ and $\lambda_2$
     of the spatial covariance matrix
     $\Sigma$, here illustrated in terms of a uniform distribution if
     isotropic receptive field shapes on a
     hemisphere, which will correspond to anisotropic affine Gaussian
     receptive fields when mapping rotationally symmetric Gaussian
     receptive fields from the tangent planes of the sphere
     to the image domain by orthographic projection.
   (right) First-order directional derivatives of the zero-order affine Gaussian kernels
   according to that distribution, with the spatial direction for
   computing the directional derivatives aligned with the orientation
   of one of the eigenvectors of the spatial covariance matrix
   $\Sigma$ that determines the shape of the affine Gaussian
   kernel. When going from the center of each of these images to the
   boundaries, the eccentricity, defined as the ratio $\epsilon =
   \sqrt{\lambda_{min}/\lambda_{max}}$ between the square roots of the smaller and
   larger eigenvalues $\lambda_{min}$ and $\lambda_{max}$
   of $\Sigma$ varies from 1 to 0, where the
   receptive fields at the center are maximally isotropic, whereas the
   receptive fields at the boundaries are maximally anisotropic.}
   \label{fig-aff-Gauss-hemisphere}
\end{figure}

\begin{figure}[hbtp]
  \begin{center}
    \begin{tabular}{c}
     \hspace{-2mm} {\footnotesize $-T(x, t;\; s, \tau)$} \hspace{-2mm} \\
      \hspace{-2mm} \includegraphics[width=0.13\textwidth]{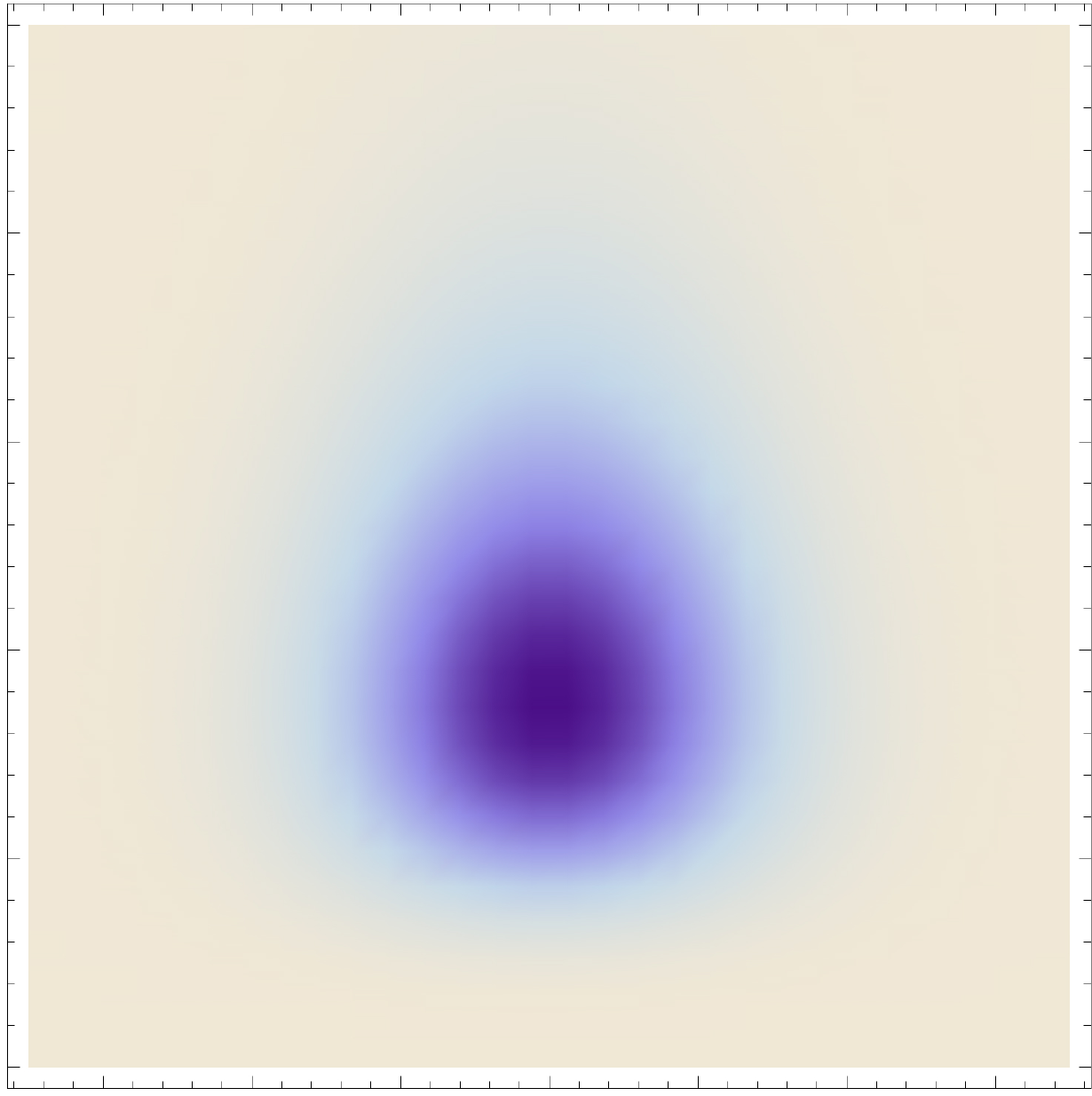} \hspace{-2mm} \\
    \end{tabular} 
  \end{center}
  \vspace{-6mm}
  \begin{center}
    \begin{tabular}{cc}
      \hspace{-2mm} {\footnotesize $T_x(x, t;\; s, \tau)$} \hspace{-2mm} 
      & \hspace{-2mm} {\footnotesize $T_t(x, t;\; s, \tau)$} \hspace{-2mm} \\
      \hspace{-2mm} \includegraphics[width=0.13\textwidth]{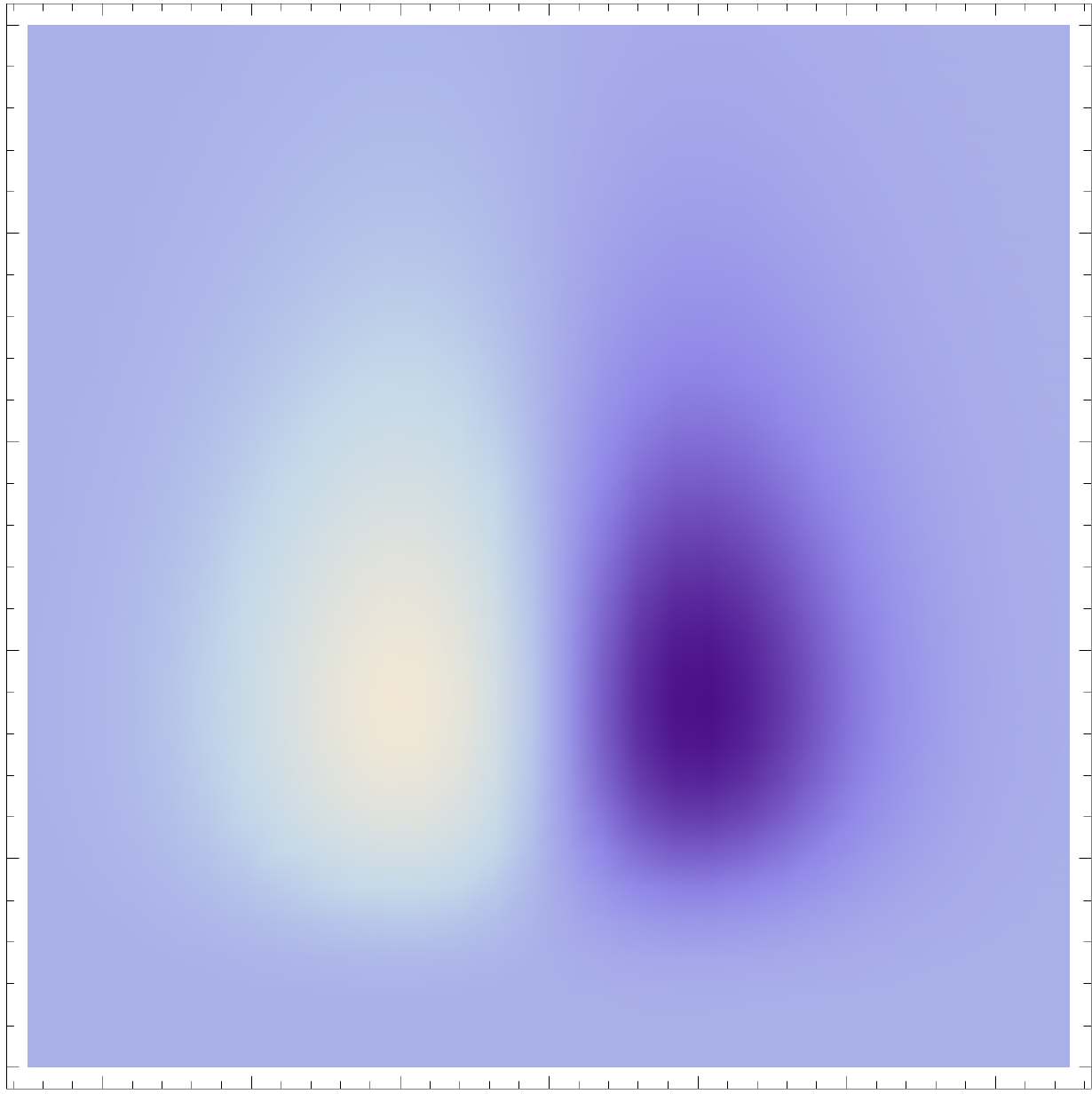} \hspace{-2mm} &
      \hspace{-2mm} \includegraphics[width=0.13\textwidth]{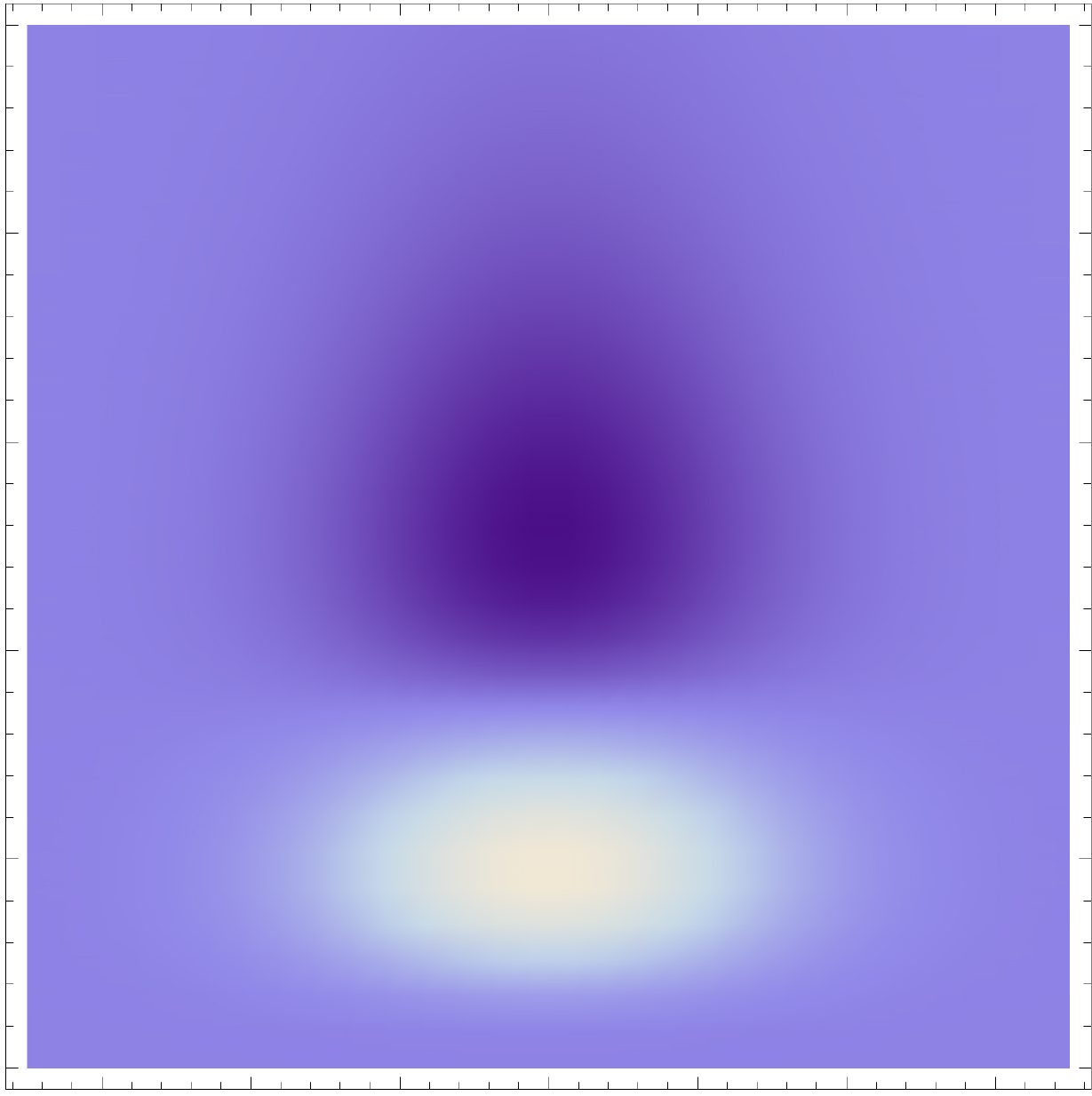} \hspace{-2mm} \\
    \end{tabular} 
  \end{center}
  \vspace{-6mm}
  \begin{center}
    \begin{tabular}{ccc}
      \hspace{-2mm} {\footnotesize $T_{xx}(x, t;\; s, \tau)$} \hspace{-2mm} 
      & \hspace{-2mm} {\footnotesize $T_{xt}(x, t;\; s, \tau)$} \hspace{-2mm} 
      & \hspace{-2mm} {\footnotesize $T_{tt}(x, t;\; s, \tau)$} \hspace{-2mm} \\
      \hspace{-2mm} \includegraphics[width=0.13\textwidth]{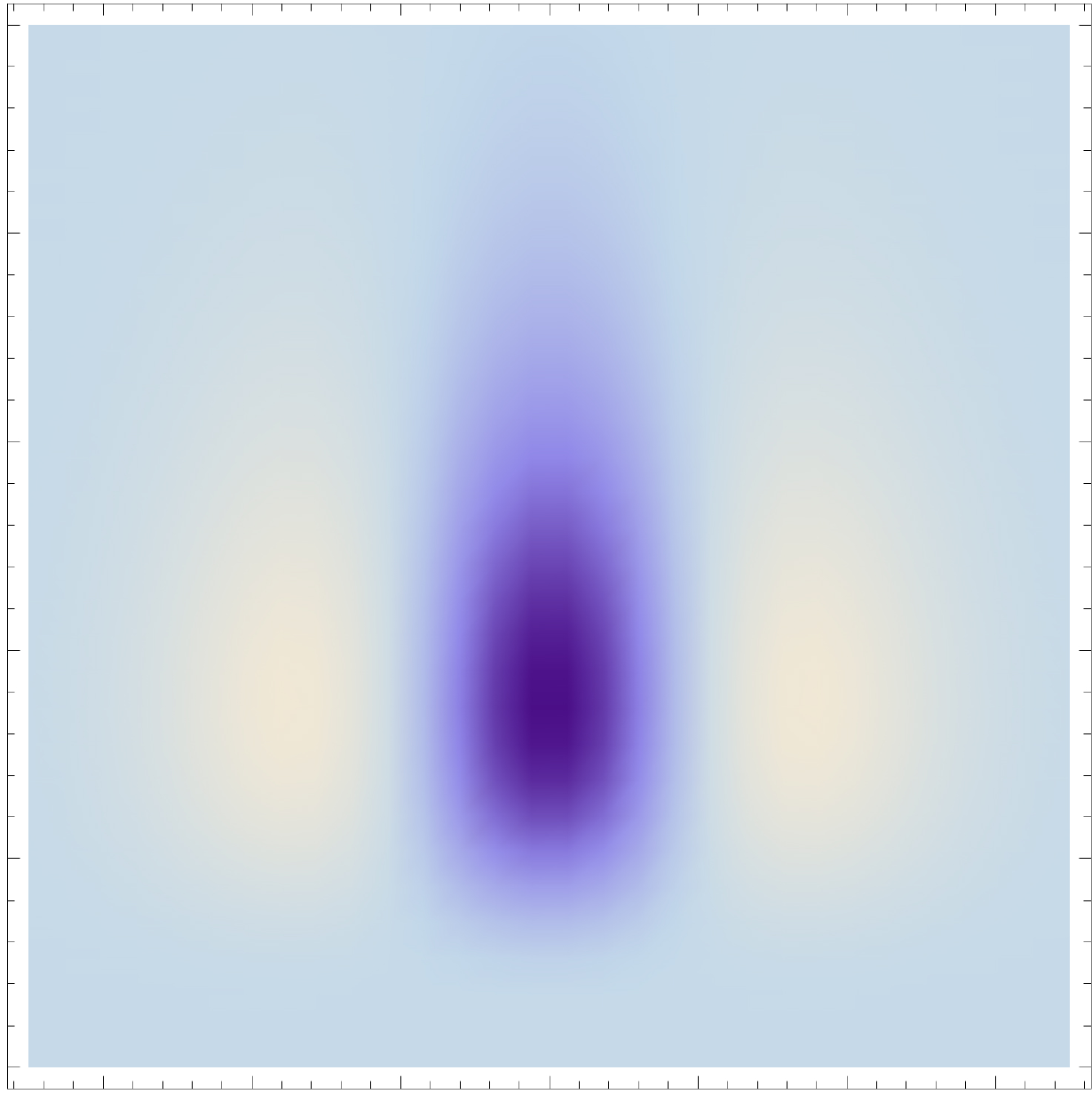} \hspace{-2mm} &
      \hspace{-2mm} \includegraphics[width=0.13\textwidth]{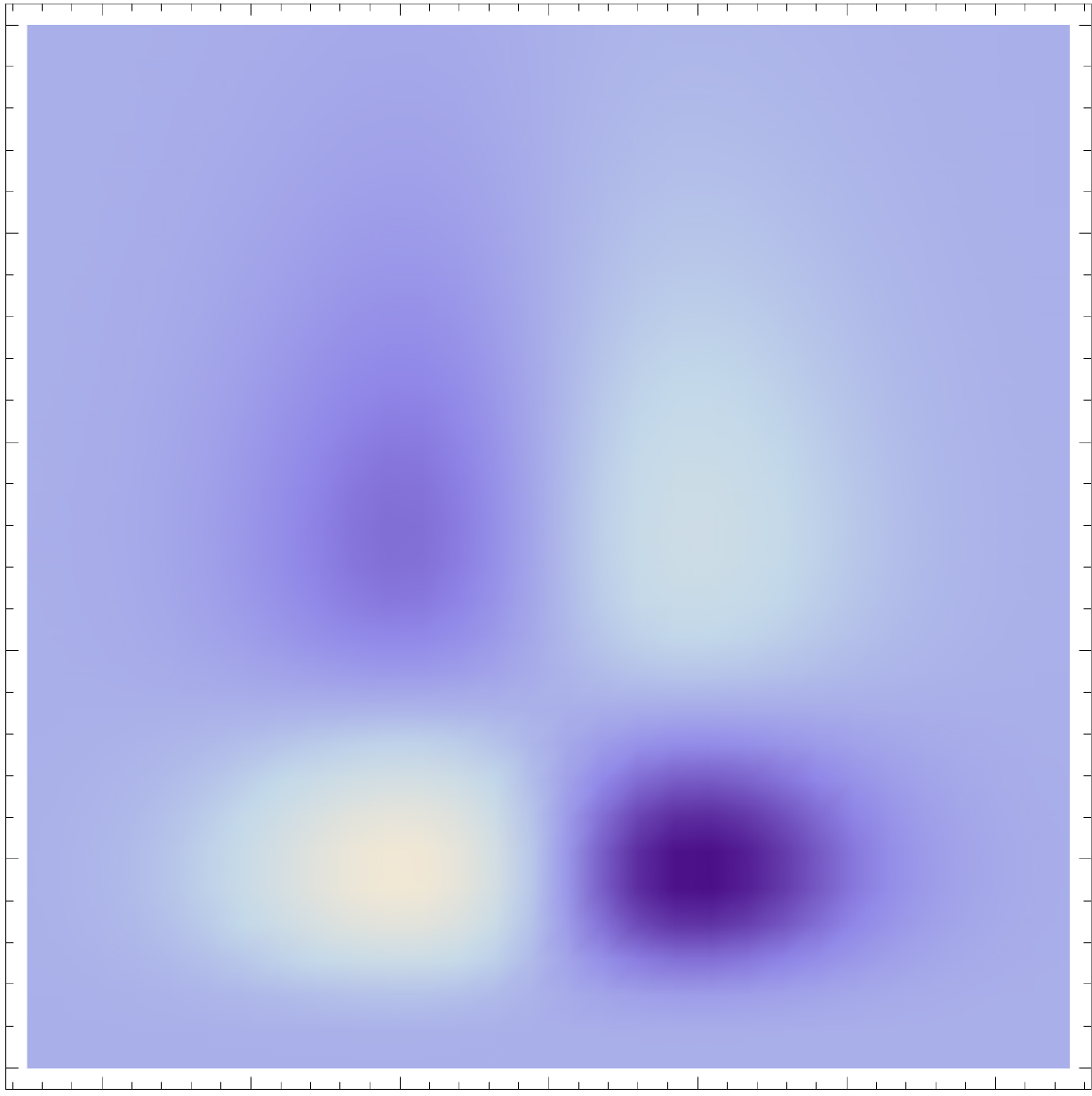} \hspace{-2mm} &
      \hspace{-2mm} \includegraphics[width=0.13\textwidth]{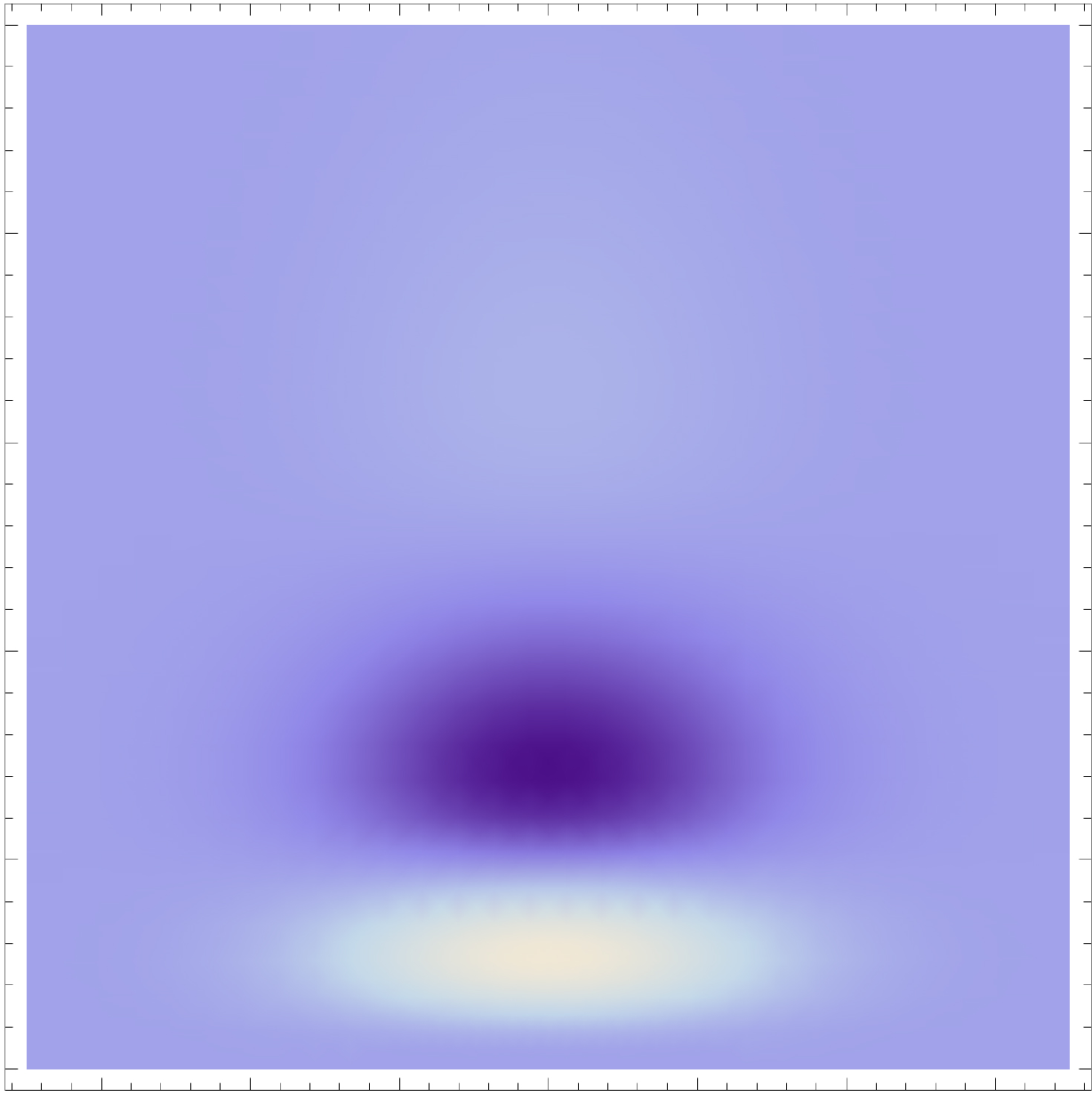} \hspace{-2mm} \\
    \end{tabular} 
  \end{center}
  \caption{Space-time separable kernels according to the time-causal
    spatio-temporal receptive field model in
    (\ref{eq-spat-temp-RF-model-der-norm-caus}) up to order two
    over a 1+1-D spatio-temporal domain for image velocity $v =
    0$, using the time-causal limit kernel
    (\ref{eq-time-caus-limit-kern}) with distribution parameter
    $c = \sqrt{2}$ for smoothing over the
    temporal domain, at spatial scale $s = 1$ and temporal scale
    $\tau = 1$. (Horizontal dimension: space $x \in [-3.5, 3.5]$.
    Vertical dimension: time $t \in [0, 5]$.)}
  \label{fig-non-caus-sep-spat-temp-rec-fields}
\end{figure}

\begin{figure}[hbtp]
  \begin{center}
    \begin{tabular}{c}
     \hspace{-2mm} {\footnotesize $-T(x, t;\; s, \tau, v)$} \hspace{-2mm} \\
      \hspace{-2mm} \includegraphics[width=0.13\textwidth]{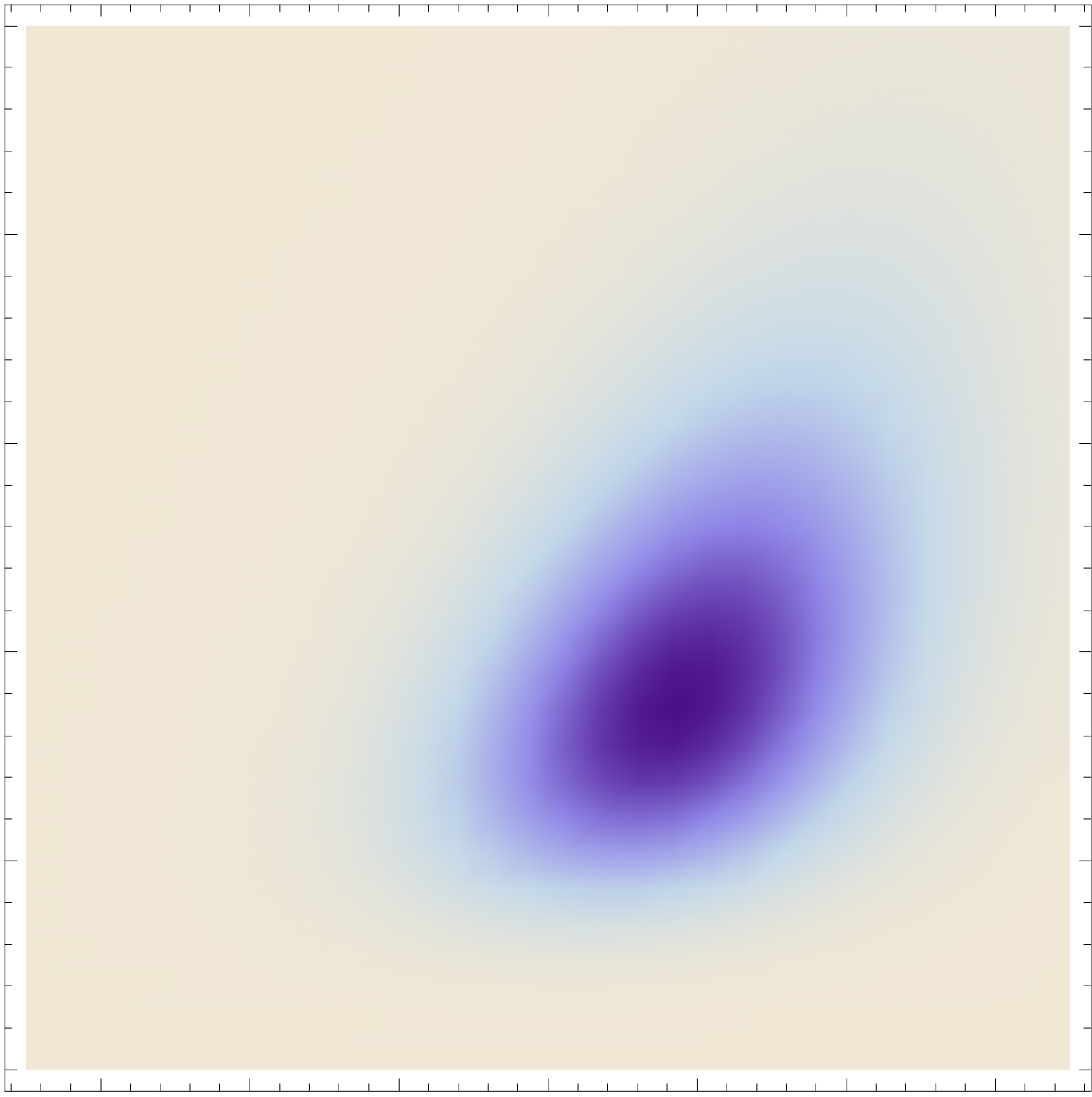} \hspace{-2mm} \\
    \end{tabular} 
  \end{center}
  \vspace{-6mm}
  \begin{center}
    \begin{tabular}{cc}
      \hspace{-2mm} {\footnotesize $T_x(x, t;\; s, \tau, v)$} \hspace{-2mm} 
      & \hspace{-2mm} {\footnotesize $T_t(x, t;\; s, \tau, v)$} \hspace{-2mm} \\
      \hspace{-2mm} \includegraphics[width=0.13\textwidth]{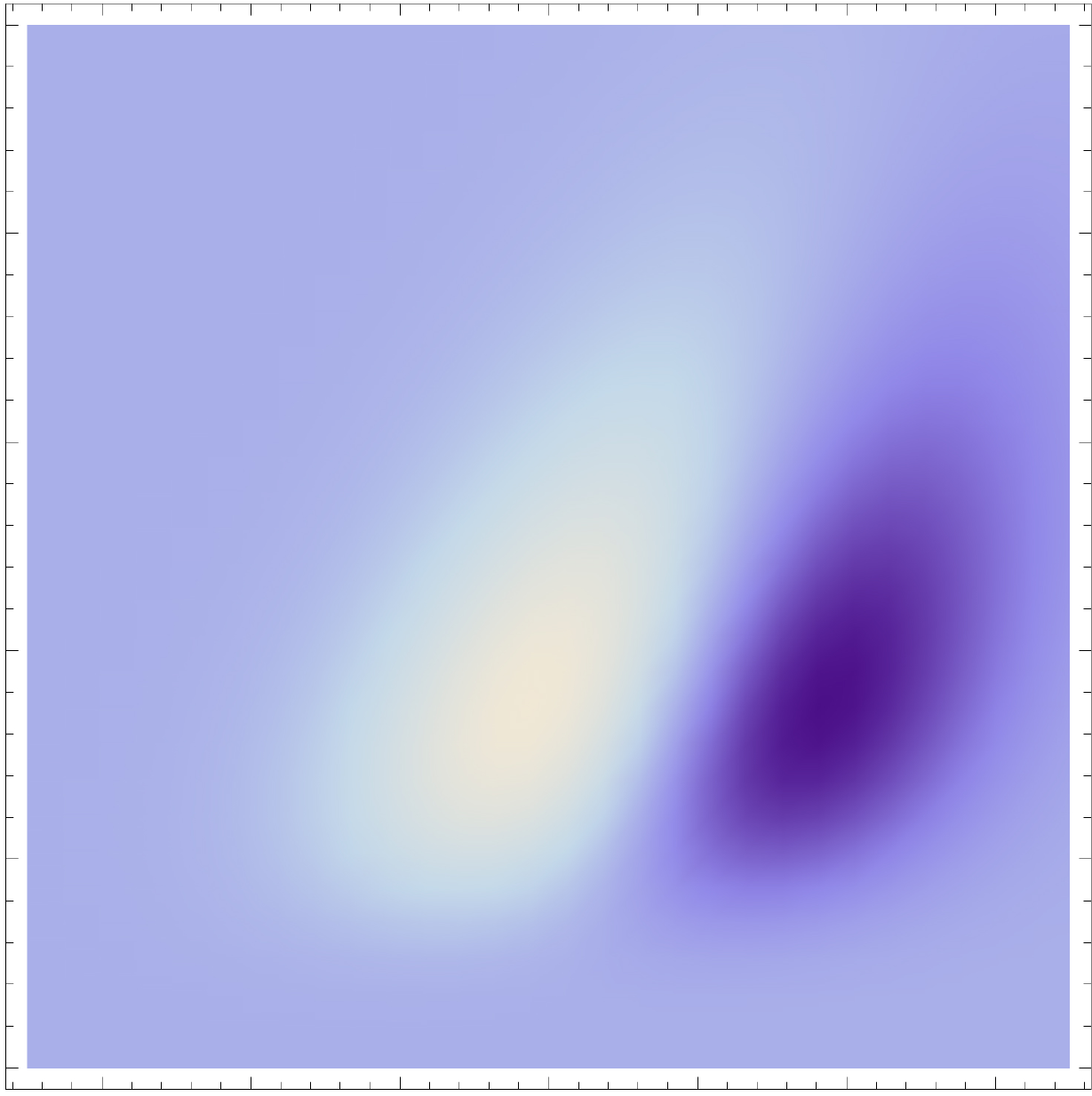} \hspace{-2mm} &
      \hspace{-2mm} \includegraphics[width=0.13\textwidth]{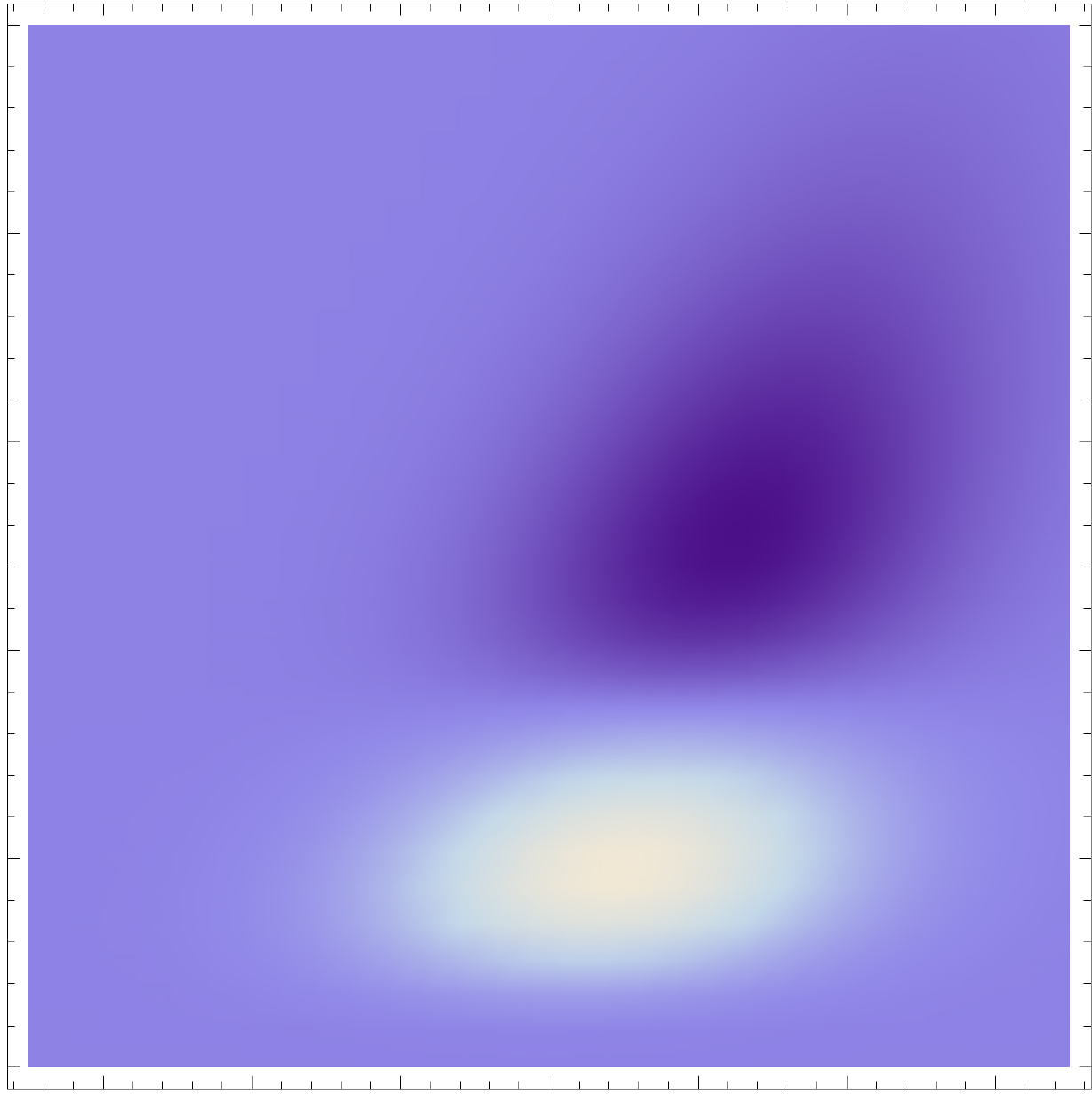} \hspace{-2mm} \\
    \end{tabular} 
  \end{center}
  \vspace{-6mm}
  \begin{center}
    \begin{tabular}{ccc}
      \hspace{-2mm} {\footnotesize $T_{xx}(x, t;\; s, \tau, v)$} \hspace{-2mm} 
      & \hspace{-2mm} {\footnotesize $T_{xt}(x, t;\; s, \tau, v)$} \hspace{-2mm} 
      & \hspace{-2mm} {\footnotesize $T_{tt}(x, t;\; s, \tau, v)$} \hspace{-2mm} \\
      \hspace{-2mm} \includegraphics[width=0.13\textwidth]{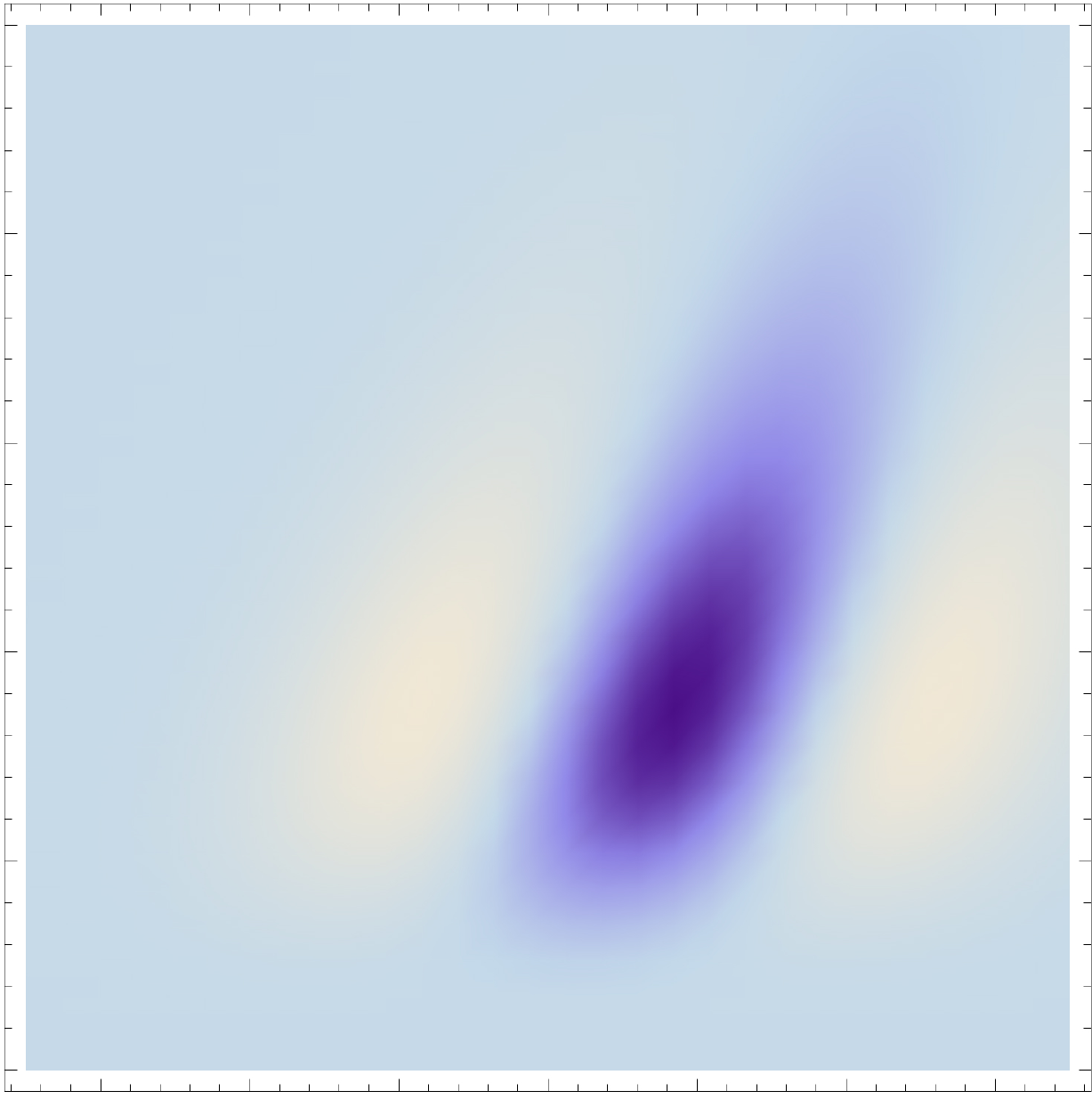} \hspace{-2mm} &
      \hspace{-2mm} \includegraphics[width=0.13\textwidth]{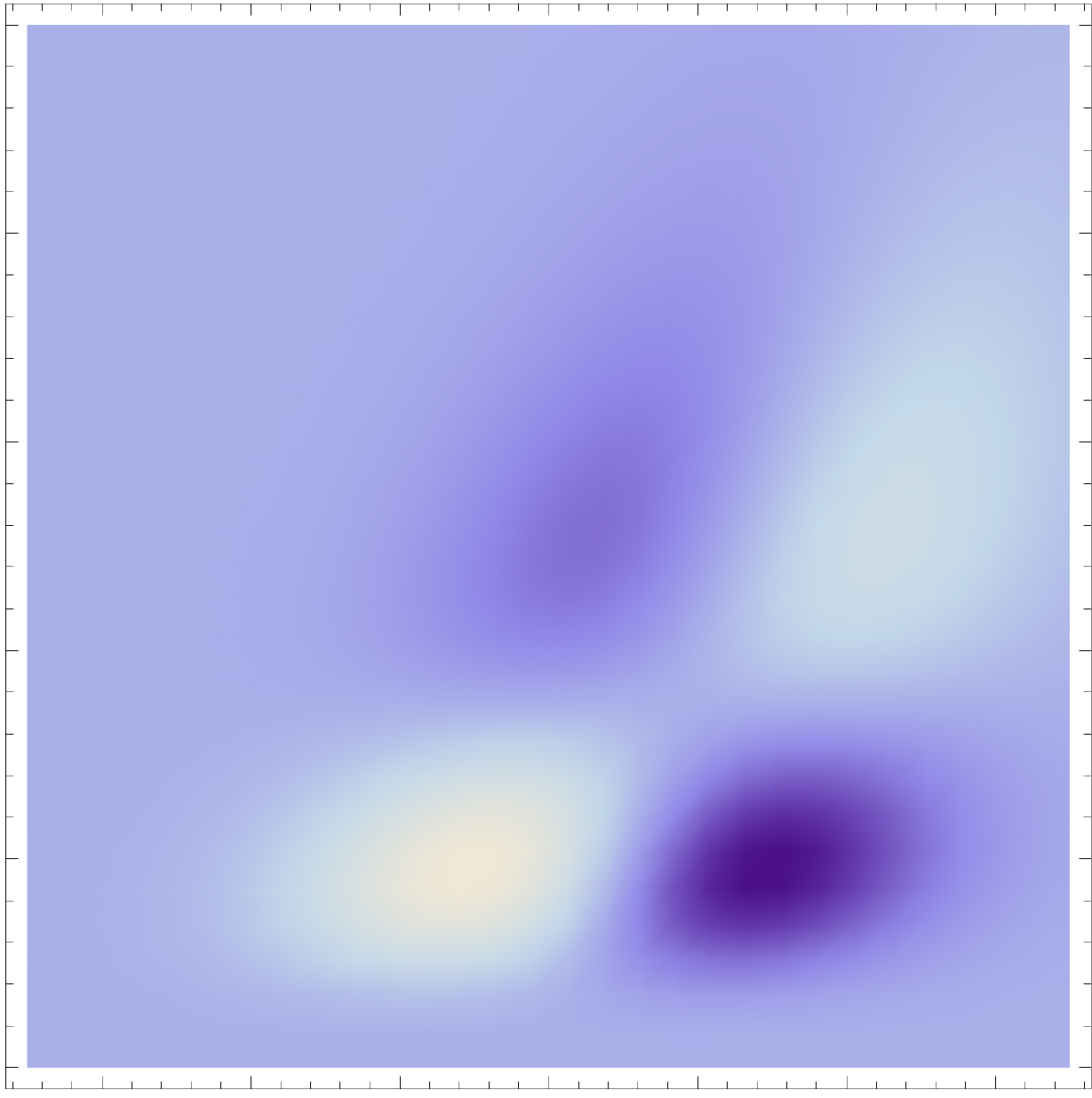} \hspace{-2mm} &
      \hspace{-2mm} \includegraphics[width=0.13\textwidth]{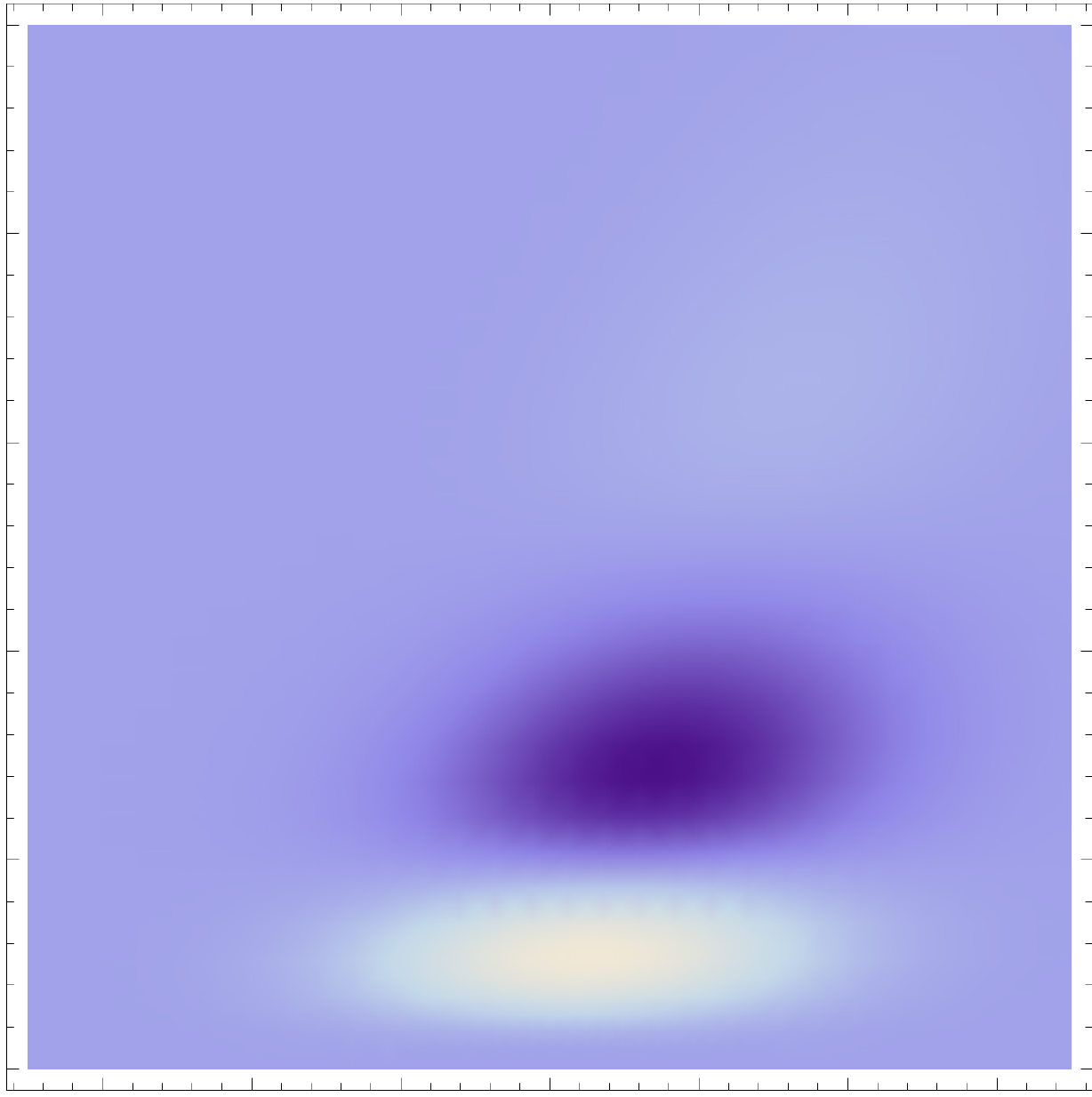} \hspace{-2mm} \\
    \end{tabular} 
  \end{center}
    \caption{Non-separable spatio-temporal kernels according to the time-causal
    spatio-temporal receptive field model in
    (\ref{eq-spat-temp-RF-model-der-norm-caus})
    up to order two
    over a 1+1-D spatio-temporal domain for image velocity $v = 1/2$,
    using the time-causal limit kernel (\ref{eq-time-caus-limit-kern})
    with distribution parameter $c = \sqrt{2}$ for smoothing over the
    temporal domain, at spatial scale $s = 1$ and temporal scale
    $\tau = 1$. (Horizontal dimension: space $x \in [-3.5, 3.5]$.
    Vertical dimension: time $t \in [0, 5]$.)}
  \label{fig-non-caus-vel-adapt-spat-temp-rec-fields}
\end{figure}

\begin{figure}[hbtp]
  \begin{center}
    \begin{tabular}{ccccccc}
      \hspace{-2mm}
      \includegraphics[width=0.12\textwidth]{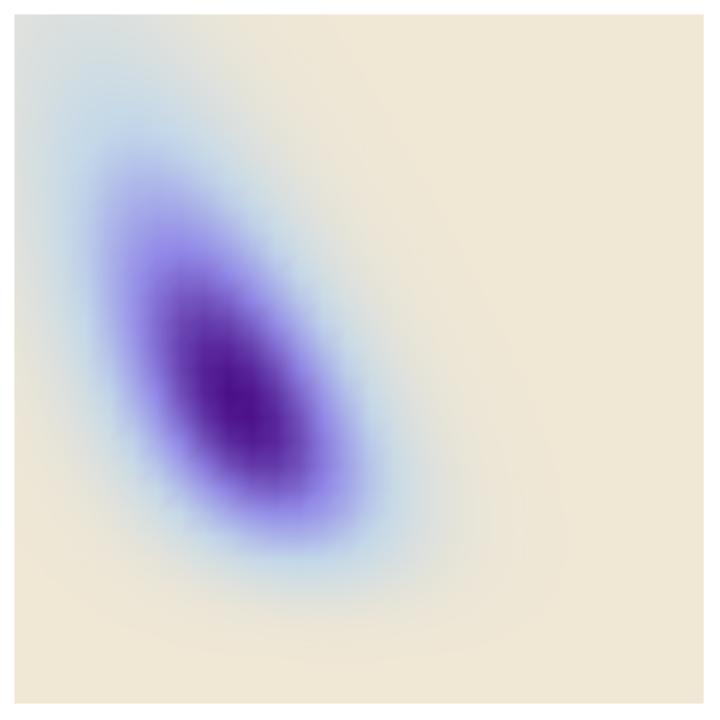} \hspace{-2mm}
      & \includegraphics[width=0.12\textwidth]{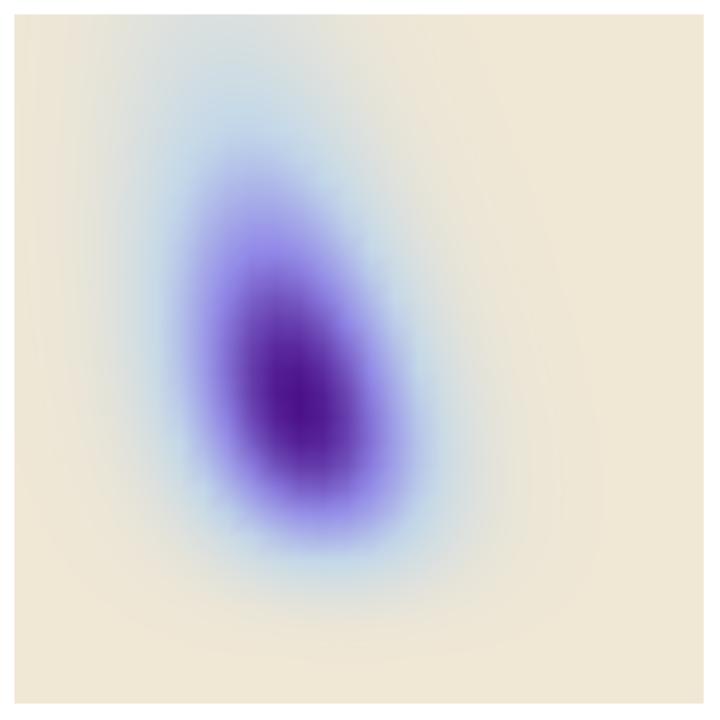} \hspace{-2mm}
      & \includegraphics[width=0.12\textwidth]{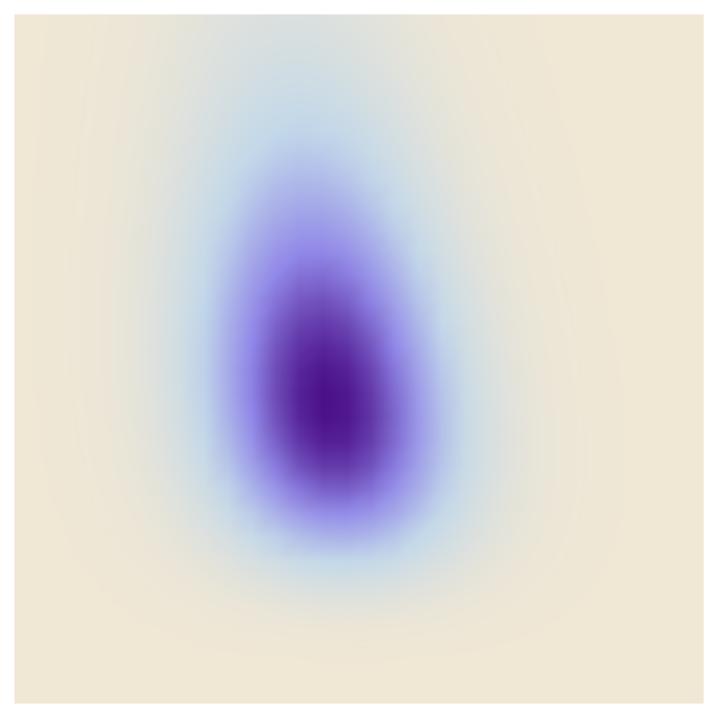} \hspace{-2mm}
      & \includegraphics[width=0.12\textwidth]{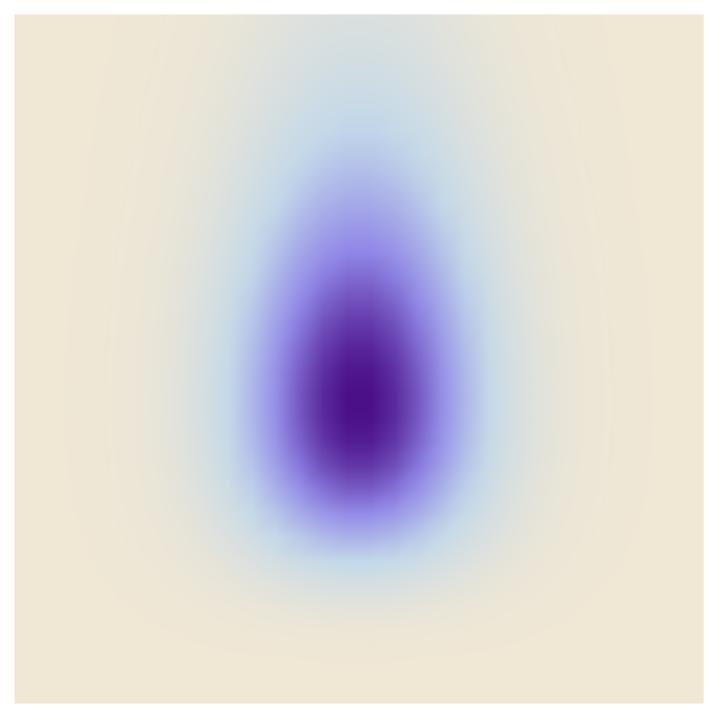} \hspace{-2mm}
      & \includegraphics[width=0.12\textwidth]{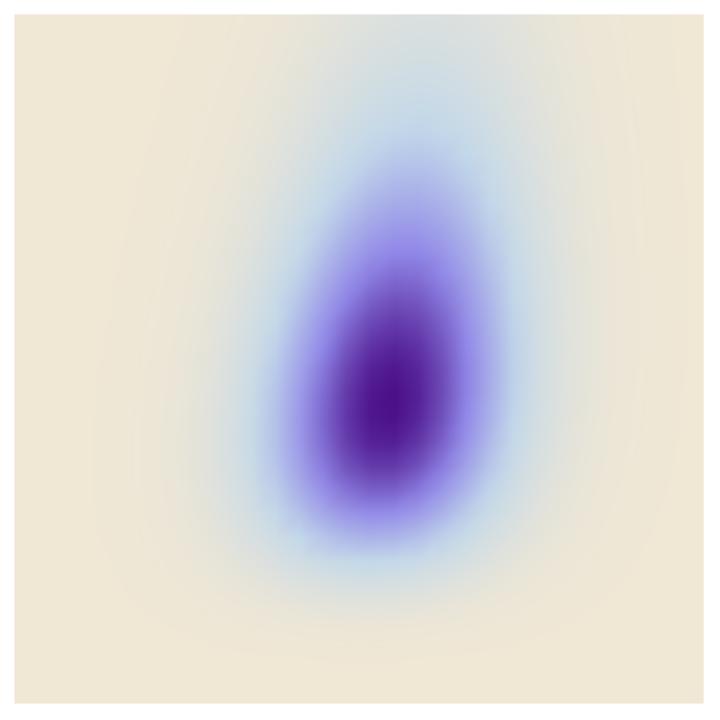} \hspace{-2mm}
      & \includegraphics[width=0.12\textwidth]{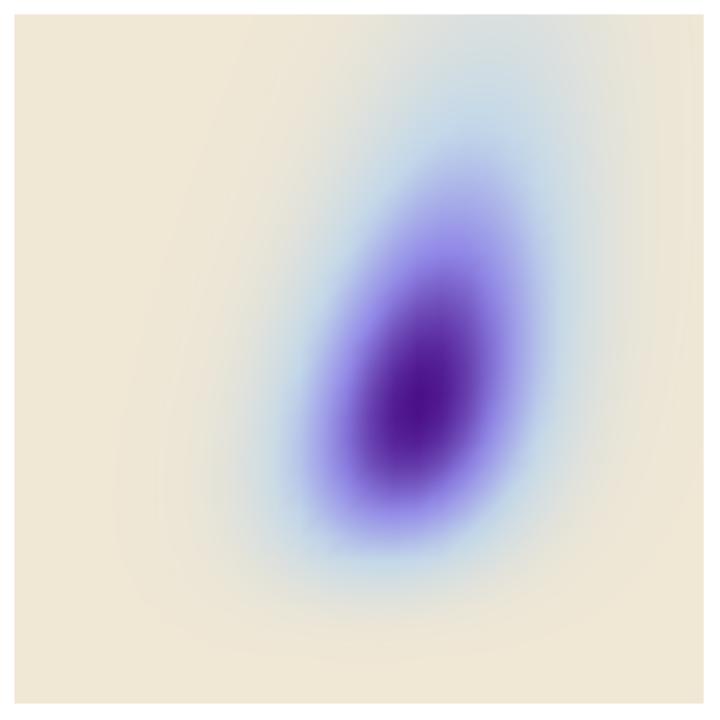} \hspace{-2mm}
      & \includegraphics[width=0.12\textwidth]{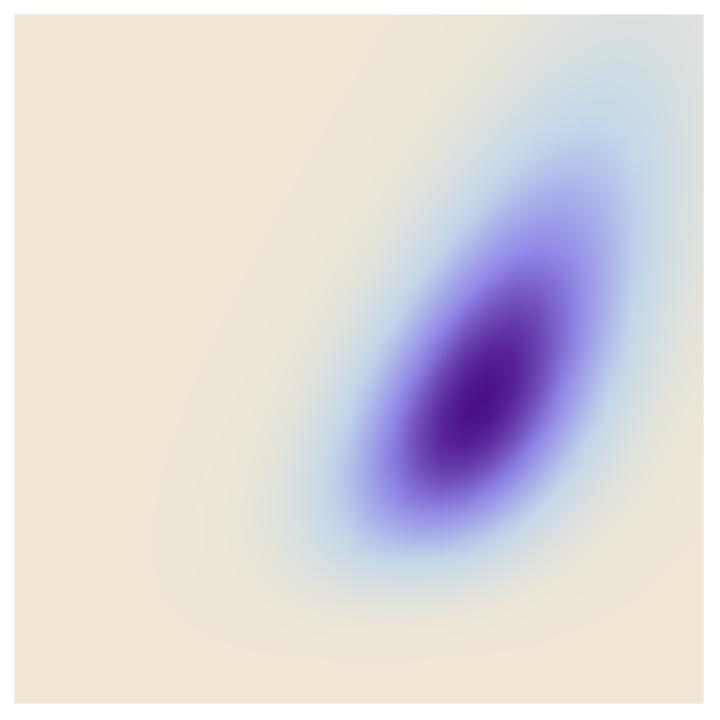} \hspace{-2mm}
    \end{tabular} 
  \end{center}
  \caption{Illustration of the variability of receptive field shapes
    over Galilean motions. This figure shows zero-order spatio-temporal receptive fields of the form
    (\ref{eq-spat-temp-RF-model-der-norm-caus}), for a
    self-similar distribution of the velocity values $v$, using Gaussian
    smoothing over the spatial domain and smoothing with the
    time-causal limit kernel (\ref{eq-time-caus-limit-kern})
    with distribution parameter $c = \sqrt{2}$ over the temporal domain, in the case of a
    1+1-D time-causal spatio-temporal domain, for spatial scale
    parameter $s = 1$ and temporal scale parameter $\tau = 4$.
    These primitive spatio-temporal smoothing operations should, in
    turn, be complemented by spatial differentiation $\partial_x^m$
    and velocity-adapted temporal differentiation operators according to
    $\partial_{\bar t}^n = (v \, \partial_x + \partial_t)^n$ for
    different orders $m$ and $n$ of spatial and temporal
    differentiation, respectively, to generate the corresponding family
    of spatio-temporal receptive fields of the form
    (\ref{eq-spat-temp-RF-model-der-norm-non-caus}).
    The central receptive
    field in this illustration, for zero image velocity is space-time
    separable, whereas the other receptive fields, for velocity values
    $v = -0.4, -0.2, -0.1, 0.1, 0.2$ and $0.4$ from left to right, are not separable.
    (Horizontal dimension: space $x \in [-4, 4]$.
    Vertical dimension: time $t \in [0, 8]$.)}
    \label{fig-distr-vel-adapt-spattemp-kernels-1+1-D}
\end{figure}

Compare with the orientation maps that would be
generated from the distribution of spatial receptive fields shown in
Figure~\ref{fig-aff-Gauss-hemisphere}, although do note that in that
illustration, the same spatial orientation is represented at two
opposite spatial positions in relation to the origin, whereas the
biological orientation maps around the pinwheels only represent a spatial orientation
once. That minor technical problem can, however, be easily fixed, by
mapping the angular representation in
Figure~\ref{fig-aff-Gauss-hemisphere} to a double angle
representation,
which would then identify
the directional derivatives of affine Gaussian kernels with opposite
polarity, in other words kernels that correspond to flipping the sign
of the affine Gaussian derivatives.

An interesting follow-up question for biological vision research does 
thus concern if it can be neurophysiologically established if the
distribution of the spatial shapes of the the spatial smoothing
component of the simple cells spans
a larger part of the affine group than a mere expansion over spatial
rotations?%
\footnote{The view followed in this treatment, that the primary visual cortex performs a substantial
  expansion of the image and video data over the parameters of the geometric image
  and video transformations, is consistent with the substantial expansion of
  measurement data that is performed from the LGN, with about 1~M
  neurons and 1~M output channels, to V1, with 190~M neurons and 37~M
  output channels, see (DiCarlo {\em et al.\/}\ \citeyear{DiCZocRus12-Neuron} Figure~3).}
What would then the distribution be over different eccentricities, {\em
  i.e\/}, different ratios%
\footnote{And taking the square root of the ratio between the
  eigenvalues of the spatial covariance matrix $\Sigma$,
to base the eccentricity measurement on measurements in unit of $[\mbox{length}]$,
as opposed to units of $[\mbox{length}^2]$. Geometrically, variations
in this ratio between the square root of the ratio between the
eigenvalue does, for example, make it possible to handle variations in the slant angle
of a local surface patch viewed under monocular projection.}
between the eigenvalues of the spatial
covariance matrix $\Sigma$, if the spatial component of each simple
cells is modelled as a directional derivative (of a suitable order) of
an affine Gaussian kernel? Note, in relation to the neurophysiological
measurements by Nauhaus {\em et al.\/}
(\citeyear{NauBenCarRin09-Neuron}),
which show a lower orientation selectivity at the pinwheels and
increasing directional selectivity when moving further away from the pinwheel,
that the spatial receptive fields based on the
maximally isotropic affine Gaussian kernels in the centers of
Figure~\ref{fig-aff-Gauss-hemisphere} (for eccentricity $\epsilon = 1$)
would have the lowest degree of
orientation selectivity, whereas the spatial receptive fields towards
the periphery (for $\epsilon$ decreasing towards 0) would have the highest degree of
orientation selectivity, see (Lindeberg \citeyear{Lin23-arXiv-OriSel})%
\footnote{In relation to the fact that the measurements of orientation
  selectivity by Nauhaus {\em et al.\/}
  (\citeyear{NauBenCarRin09-Neuron}) may most likely involve complex
  cells, whereas the predictions in this theory are based on the
  response properties of simple cells, it can be noted that that this
  qualitative result concerning the orientation selectivity of simple
  cells, based on the affine Gaussian derivative model,
  will also extend to complex cells, if the complex cells can be
  modelled as operating on the output of simple cells as done in
  (Lindeberg \citeyear{Lin20-JMIV} Section~5),
  see (Lindeberg \citeyear{Lin23-arXiv-OriSel}) for a detailed
  analysis of this subject.}
for an in-depth treatment of this topic.

If an expansion over eccentricities for simple cells
could be established over the spatial domain,
it would additionally be interesting to investigate if such an expansion would be
coupled also to the expansion over image velocities $v$ in space-time%
\footnote{We implicitly regard variability over image
velocities as established, since velocity-tuned receptive fields have been recorded
for different image velocities.}
for joint spatio-temporal receptive fields, or if
a potential spatial expansion over eccentricities in the affine group
(alternatively over some other subset of the affine group) is decoupled from
the image velocities in the Galilean group. If those expansions are
coupled, then the dimensionality of the parameter space would be
substantially larger, while the dimensionality over two separate
expansions over eccentricities {\em vs.\/} motion directions would be
substantially lower. Is it feasible for the earliest layers of
receptive field to efficiently represent those dimensions of the
parameter space jointly, which would then
enable higher potential accuracy in the derivation of cues to local
surface orientation and surface shape from moving objects
that are not fixated by the observer, and thus moving with non-zero
image velocity relative to the viewing direction.
Or do efficiency arguments call for a separation of those dimensions
of the parameter space, into separate spatial and motion pathways, so that
accurate surface orientation and shape estimation can only be
performed with respect to viewing directions that are fixated by the observer in
relation to a moving object?

It would additionally be highly interesting to characterize to what extent
the early stages in the visual system perform expansions of receptive
fields over
multiple spatial and temporal scales. In the retina, the spatial
receptive fields mainly capture a lowest spatial scale level, which
increases linearily with the distance from the fovea, see
(
 Lindeberg \citeyear{Lin13-BICY} Section~7).
Experimental evidence do on the other hand demonstrate that biological
vision achieves scale invariance over wide ranges of scale
(Biederman and Cooper \citeyear{BieCoo92-ExpPhys};
 Logothetis {\em et   al.\/} \citeyear{LogPauPog95-CurrBiol};
 Ito {\em et al.\/} \citeyear{ItoTamFujTan95-JNeuroPhys};
 Furmanski and Engel \citeyear{FurEng00-VisRes};
 Hung {\em et al.\/} \citeyear{HunKrePogDiC05-Science};
 Isik {\em et al.\/} \citeyear{IsiMeyLeiPog13-JNPhys}).
Using the semi-group properties
of the rotationally symmetric as well as the affine Gaussian kernels%
\footnote{The semi-group property over spatial scales $s$ of the rotationally symmetric
  Gaussian kernel $g(x;\; s)$ implies that the convolution of two
  rotationally symmetric Gaussian
kernels with each other is also a rotationally symmetric Gaussian
kernel, and with added scale
parameters $g(x;\; s_1) * g(x;\; s_2) = g(x;\; s_1+s_2)$, whereas the semi-group
property over the spatial covariance matrices $\Sigma$
of the affine Gaussian kernel means that the convolution of
two affine Gaussian kernels is also an affine Gaussian kernel, and
with added spatial covariance matrices
$g(x;\; \Sigma_1) * g(x;\; \Sigma_2) = g(x;\;
\Sigma_1+\Sigma_2)$. These semi-group properties do, in turn, mean that
spatial smoothed representations with the rotationally symmetric
Gaussian kernel are related according to the spatial cascade smoothing
property $L(x;\; s_2) = g(x;\; s_2 - s_1)* L(x;\; s_1)$, provided that
$s_2 > s_1$, whereas spatially smoothed representations with the
affine Gaussian kernel are related according to spatial cascade smoothing property
$L(x;\; \Sigma_2) = g(x;\; \Sigma_2 - \Sigma_1)* L(x;\; \Sigma_1)$,
provided that $\Sigma_2 - \Sigma_1$ is a symmetric positive definite matrix.
With regard to the spatial component of the model
for the receptive fields of LGN neurons (\ref{eq-lgn-model-spat}),
the spatial cascade smoothing property originating from the semi-group
property of the rotationally symmetric Gaussian kernel implies that the
output from the spatial LGN model (\ref{eq-lgn-model-spat}) at spatial
scale $s_2$ is related to the corresponding output at spatial scale $s_1$
according to $L_{LGN}(x;\; s_2) = s^{s_2 - s_1} g(x;\; s_2 - s_1)* L_{LGN}(x;\; s_1)$,
provided that $s_2 > s_1$, whereas the output from the
spatial model of the receptive fields of simple cells
(\ref{eq-spat-RF-model}) at spatial scale $s_2$ is related to the
corresponding output at spatial scale $s_1$ according to
$L_{simple}(x;\; s_2, \Sigma) =
s^{(s_2 - s_1)m/2} g(x;\; (s_2 - s_1)\Sigma)* L_{simple}(x;\; s_1, \Sigma)$,
provided that $s_2 > s_1$ and that the direction of the directional derivative operator in
this model for simple cells is aligned to the eigenvectors of the spatial covariance
matrix $\Sigma$.},
it is in
principle possible to compute coarser scale representations from finer
scale levels by adding complementary spatial smoothing stages in
cascade. The time-causal limit kernel (\ref{eq-time-caus-limit-kern})
used for temporal smoothing in
our spatio-temporal receptive field model
does also obey a cascade smoothing property over temporal scales,
which makes it possible
to compute representations at coarser temporal scales from
representations at finer spatial scales, by complementary
(time-causal) temporal filtering, in terms of first-order temporal
integrators coupled in cascade%
\footnote{Due to the cascade smoothing property of the time-causal
  limit kernel over temporal scales, the time-causal limit kernels $\Psi(t;\; \tau, c)$ at
  adjacent temporal scales are related according to
  $\Psi(t;\; \tau, c)
  = h_{exp}(t;\; \tfrac{\sqrt{c^2-1}}{c} \sqrt{\tau}) * \Psi(t;\; \tfrac{\tau}{c^2}, c)$,
  with $h_{exp}(t;\; \mu)$ denoting a truncated exponential kernel
  with time constant $\mu$ according to
  $h_{exp}(t;\; \mu) = \frac{1}{\mu_k} e^{-t/\mu_k}$ and $c$ being the
  distribution parameter of the time-causal limit kernel
  (Lindeberg \citeyear{Lin23-BICY} Equation~(28)).
  This does, in turn, mean that temporal derivatives of purely temporally smoothed
  representations at adjacent temporal scales are related according to
  $L_{t^n}(t;\; \tau, c) 
  = h_{exp}(t;\; \tfrac{\sqrt{c^2-1}}{c} \sqrt{\tau}) * L_{t^n}(t;\; \tfrac{\tau}{c^2}, c)$.}
(Lindeberg \citeyear{Lin23-BICY}).
Do the earliest layers in a biological visual system
explicitly represent the image and video data by expansions of
receptive fields over multiple
spatial and temporal scales, or do the earliest stages in the vision
system instead only represent a lowest range of spatial and temporal scales
explicitly, to then handle coarser spatial and temporal scales by other
mechanisms?

The generalized Gaussian derivative model for visual receptive fields
can finally be seen as biologically plausible, in the respect that the
computations needed to perform the underlying spatial smoothing
operations and spatial derivative
computations can be performed by local spatial computations.
The spatial smoothing is modelled by diffusion equations
(\ref{eq-vel-adapt-scsp-diff-eq-spat}), which can
be implemented by local computations based on connections between
neighbours, and thus be performed
by groups of neurons that interact with each other spatially by local connections. Spatial
derivatives can also be approximated by local nearest neighbour
computations. The temporal smoothing operation in the time-causal model 
based on smoothing with the time-causal limit kernel
corresponds to first-order temporal integrators coupled in cascade, with
very close relations to our understanding of temporal processing in neurons. Temporal
derivatives can, in turn, be computed from linear combinations of
temporally smoothed scale channels over multiple temporal scales,
based on a similar recurrence relation over increasing orders of
temporal differentiation as in
(Lindeberg and Fagerstr{\"o}m \citeyear{LF96-ECCV}, Equation~(18)).
Thus, all the individual computational primitives needed to implement
the spatial as well as the spatio-temporal receptive fields according
to the generalized Gaussian derivative model can, in principle, be
performed by computational mechanisms available to a network of biological neurons.

\subsubsection{Testable hypotheses from the theoretical predictions}

For a visual neuron in the primary visual cortex that can be well modelled as a
simple cell, let us assume that the spatio-temporal dependency of its
receptive field can be modelled as a combination of a dependency on a
spatial function $h_{space}(x_1, x_2)$ and a dependency on a
temporal function $h_{time}(t)$ of the joint form%
\footnote{This model is inspired by the theoretically derived spatio-temporal
  receptive field model (\ref{eq-spat-temp-RF-model-der-norm-caus}),
  however, generalized to more spatial and
  temporal dependency functions than directional derivatives of affine Gaussian kernels or
  temporal derivatives of the time-causal limit kernel, and with the
  explicit dependencies on the transformation parameters now included
  in the separate spatial and temporal dependency functions for an
  individual neuron. The model is
  also inspired by biological results concerning a preference of
  biological receptive fields to specific motion directions space-time.}
\begin{equation}
  h_{simple}(x_1, x_2, t) = h_{space}(x_1 - v_1 t, x_2 - v_2 t) \, h_{time}(t)
\end{equation}
for some value of a velocity vector $v = (v_1, v_2)$, where we here in
this model
assume that both the spatial and the temporal dependency functions
$h_{space}(x_1, x_2)$ and $h_{time}(t)$ as well as the velocity vector
$v$ are individual for each visual neuron.
Then, we can formulate testable hypotheses for the above theoretical
predictions in the following ways:

{\bf Hypothesis~1} (Expansion of spatial receptive field shapes over a larger part of
the affine group than mere rotations or uniform scale changes):
Define characteristic lengths $\sigma_{\varphi}$ and
$\sigma_{\orth \varphi}$ that measure the spatial extent of the
spatial component $h_{space}(x_1, x_2)$ of the receptive field
in the direction $\varphi$ representing the orientation of the
oriented simple cell as well as in its orthogonal direction
$\orth \varphi$, respectively. Then, if the hypothesis about an
expansion over a larger part of the affine group than mere rotations
is true, there should be a variability in the ratio of these
characteristic lengths between different neurons
\begin{equation}
  \epsilon = \frac{\sigma_{\varphi}}{\sigma_{\orth \varphi}}.
\end{equation}
If the spatial component of the receptive field $h_{space}(x_1, x_2)$
can additionally be well modelled as a directional derivative of an affine Gaussian
kernel for some order of spatial differentiation,
then this ratio between the characteristic lengths
can be taken as the ratio between the effective scale parameters%
\footnote{Note that the directional derivative of an affine Gaussian
  kernel will be separable in a coordinate system aligned to the
  orientation of the receptive field, provided that the direction for
  computing the directional derivative is aligned with one of the
  eigendirections of the covariance matrix $\Sigma$ associated with
  the affine Gaussian kernel.}
of the affine Gaussian kernel in the directions of the orientation of the
receptive field and its orthogonal direction
\begin{equation}
  \epsilon = \sqrt{\frac{s_{\varphi}}{s_{\orth \varphi}}}.
\end{equation}

{\bf Hypothesis~2} (Joint expansion over image velocities and spatial
eccentricities of the spatio-temporal receptive fields):
Assuming that there is a variability in the ratio $\epsilon$ 
between the orthogonal characteristic lengths of the spatial components
of the receptive fields between different neurons, as well as also a
variability of the absolute velocity values 
\begin{equation}
  v_{speed} = \sqrt{v_1^2 + v_2^2},
\end{equation}
between different neurons, then these variabilities should together span a
2-D region in the composed 2-D parameter space, and not be restricted
to only a set of 1-D subspaces, which could then be seen as different
populations in a joint scatter diagram or a histogram over the
characteristic length ratio $\epsilon$
and the absolute velocity value $v_{speed}$.

{\bf Hypothesis~3} (Separate expansions over image velocities and spatial
eccentricities of the spatio-temporal receptive fields):
The neurons that show a variability over the velocity
values $v_{speed}$ all have a ratio $\epsilon$ of the characteristic
lengths in two orthogonal spatial directions for the purely spatial
component of the receptive field that is close to
constant, or confined within a narrow range.

Note that Hypotheses~2 and~3 are mutually exclusive.

\subsubsection{Quantitative characterizations of distributions of
  receptive field parameters}

In addition to investigating if the above working hypotheses hold,
which would then give more detailed insights into how the spatial and
spatio-temporal receptive field shapes of simple cells are expanded
over the degrees of freedom of the geometric image transformations,
it would additionally be
highly interesting to characterize the distributions of the
corresponding receptive field parameters, in other words, the distributions of
\begin{itemize}
\item
  the spatial characteristic
  length ratio $\epsilon$,
\item
  the image velocity $v_{speed}$,
 \item
  a typical spatial size parameter $\sigma_{space}$, which for a spatial
  dependency function $h_{space}(x_1, x_2)$ in
  (\ref{eq-spat-temp-RF-model-der-norm-caus}) corresponding to
a directional derivative of an affine Gaussian kernel for a specific
normalization of the spatial covariance matrix $\Sigma$ would
correspond to the square root of the spatial scale parameter,
{\em i.e.\/}, $\sigma_{space} = \sqrt{s}$, and
\item
  a typical temporal duration parameter $\sigma_{time}$, which for the
  temporal dependency function $h_{time}(t)$ in
  (\ref{eq-spat-temp-RF-model-der-norm-caus}) corresponding to a
  temporal derivative of the time-causal limit kernel would correspond
  to the square root of the temporal scale parameter, {\em i.e.\/},
  $\sigma_{time} = \sqrt{\tau}$.
\end{itemize}
Characterizing the distributions of these receptive field parameters,
would give quantitative measures on how well the variabilities of the
receptive field shapes of biological simple cells span the studied classes of
natural image transformations, in terms of spatial scaling
transformations, spatial affine transformations, Galilean
transformations and temporal scaling transformations.

When performing a characterization of these distributions, as well as
investigations of the above testable hypotheses, it should, however, be noted
that special care may be needed, to only pool statistics over biological
receptive fields that have a similar qualitative shape, in terms of
the number of dominant positive and negative lobes over space and time.
For the spatial and spatio-temporal receptive fields according to the
studied generalized Gaussian derivative model, this would imply
initially only collecting statistics for receptive fields that
correspond to the same orders of spatial and temporal differentiation.
Alternatively, if a
unified model can be expressed for receptive fields of differing
qualitative shape, as it would be possible if the biological receptive fields
can be well modelled by
the generalized Gaussian derivative model, statistics could also be pooled
over different orders of spatial and/or temporal differentiation, by defining the 
characteristic spatial lengths from the spatial scale
parameter $s$ according to $\sigma_{space} = \sqrt{s}$,
and the characteristic temporal durations
from the temporal scale parameter $\tau$ according to
$\sigma_{time} = \sqrt{\tau}$.

\section{Summary and discussion}

We have presented a generalized Gaussian derivative model for
modelling visual receptive fields that can be modelled as linear, and
which can be derived in an axiomatic principled manner from symmetry
properties of the environment, in combination with structural
constraints on the first stages of the visual system, to guarantee
internally consistent visual representations over multiple spatial and
temporal scales.

In a companion work (Lindeberg \citeyear{Lin21-Heliyon}), it has been
demonstrated that specific instances of receptive field models obtained within this
general family of visual receptive fields do very well model
properties of LGN neurons and simple cells in the primary visual
cortex, as established in
neurophysiological cell recordings by
DeAngelis {\em et al.\/}\ (\citeyear{DeAngOhzFre95-TINS,deAngAnz04-VisNeuroSci}),
Conway and Livingstone (\citeyear{ConLiv06-JNeurSci}) and
Johnson {\em et al.\/}\ (\citeyear{JohHawSha08-JNeuroSci}).
Indeed, based on the generalized Gaussian derivative model for visual
receptive fields, it is possible to reproduce the qualitative shape of
all the main types of receptive fields reported in these
neurophysiological studies,
including space-time separable neurons in the LGN,
and simple cells in the primary visual cortex with
oriented receptive fields having strong orientation preference over the
spatial domain, as well as being either space-time separable over the joint
spatio-temporal domain, or with preference to specific motion
directions in space-time.

Specifically, we have in this
paper focused on the transformation properties of the receptive fields
in the generalized Gaussian derivative model under geometric image
transformations, as modelled by local
linearizations of the geometric transformations between single or
multiple views of (possibly moving) objects or events in the world, expressed
in terms of spatial scaling transformations, spatial affine transformations and Galilean
transformations, as well as temporal scaling transformations. We have
shown that the receptive fields in the generalized Gaussian derivative model possess true
covariance properties under these classes of natural image
transformations. The covariance properties do, in turn,
imply that a vision system, based on populations of these
receptive fields, will have the ability to, up to first order of
approximation, handle: (i)~the perspective mapping from objects or events in the
world to the image or video domain, (ii)~handle the image deformations
induced by viewing image patterns of smooth surfaces in the world from
multiple views, as well as (iii)~the video patterns arising from viewing
objects and events in the world, that move with different velocities
relative to the observer, or (iv)~spatio-temporal events that occur
faster or slower in the world.

We argue that it is essential for a vision system to obey,
alternatively sufficiently well approximate such covariance
properties, in order to robustly be able to handle the huge
variability of image of video data generated under the influence of
natural image transformations. Based on covariant image and video
measurements at early stages
in the visual hierarchy, invariant representations can, in turn, be
computed at higher levels. 
If the early stages in the visual system
would not respect such basic covariance properties, or sufficiently
good approximations thereof, the subsequent visual computations at
higher levels would suffer from inherent measurement errors, caused by
the non-infinitesimal extent and duration of the receptive fields over
space and time, that may be otherwise hard to recover from. We do therefore
argue that the covariance properties treated in this article are
essential for both (i)~the study and modelling of biological vision and (ii)~the construction
of artificial computer vision systems. Specifically, we argue that the
influence of natural image transformations on the measurements of
local image and video information based on visual receptive fields,
is essential for understanding both the possibilities for and the
computational functions in visual perception.

\subsection{Relations to other sources of variability in image and
  video data}

Beyond the variability due to natural geometric image transformations, which are
handled by expanding the receptive field shapes over the degrees of
freedom of the  corresponding image transformations, it should
be remarked that the presented model for visual receptive fields,
does additionally have the ability to handle also other sources of
variability in image and video data.

Concerning illumination variations, it can be shown that if receptive
fields according to the
generalized Gaussian derivative model for visual receptive fields,
based on spatial and temporal derivatives of spatial and
temporal smoothing kernels, are applied to image intensities
expressed on a logarithmic brightness scale,
then the receptive field responses will be automatically
invariant to multiplicative intensity transformations,
and thus be able to handle both (i)~multiplicative changes
in the illumination and (ii)~multiplicative changes in
exposure control mechanisms.
In this way, a large source of variability regarding illumination
changes is implicitly handled by the presented theory
(Lindeberg \citeyear{Lin13-BICY} Section 2.3; 
Lindeberg \citeyear{Lin21-Heliyon} Section 3.4).

Concerning image and video noise, very fine scale receptive fields
will have the property that they may respond primarily to 
fine scale surface textures and noise, whereas the noise and the fine scale
textures will become effectively suppressed
in coarser scale receptive fields. In this way, the coarser scale
receptive fields will be more robust to image noise as well as the
influence of very fine scale surface textures.

Concerning transparencies, an interesting property of a
multi-parameter model for receptive fields, as considered
here, is also that it can respond to qualitatively
different types of image structures for different values of
the receptive field parameters. Beyond different types of
responses at different scales, which as previously considered
can handle spatial structures at different scales, by varying
the velocity parameter $v$ in the spatio-temporal receptive field model
in Equation~(\ref{eq-spat-temp-RF-model}),
such a model will have the ability to handle aspects of transparent motion, in the sense
that by considering the receptive field responses for the
parameters of Galilean motion that describe the motion
of foreground image structures {\em vs.\/}\ the background image
structure in a two-layer transparent motion. By extending the
receptive field model to binocular receptive fields over a disparity
parameter, obtained by varying the parameter $\delta_{(x_1, x_2)}$
in Equation~(\ref{eq-vel-adapt-scsp-diff-eq-spat}), such an extended
model of visual receptive fields would also have the ability to handle
static transparencies.

In these ways, the multi-parameter model of receptive fields
considered here can also serve other purposes, beyond handling geometric
image transformations, as a basis for early vision.
If aiming at extending the model to other sources of variability, such
as handling occlusions, then a
natural starting point to use is to start from the generalized
diffusion equations (\ref{eq-vel-adapt-scsp-diff-eq-spat}) and
(\ref{eq-vel-adapt-scsp-diff-eq-spat-temp})
that generate the corresponding visual receptive
fields, and then complement with explicit learning mechanisms over the
corresponding parameters in the receptive fields, and also to, for example,
complement with explicit end-stopping mechanisms to prevent the smoothing process from
extending over object boundaries.

\subsection{Relations to previous work}
\label{sec-rel-prev-work}

In their ground-breaking work, Hubel and Wiesel (\citeyear{HubWie59-Phys,HubWie62-Phys,HubWie68-JPhys,HubWie05-book}) characterized properties of the receptive fields for simple and complex cells in the primary visual cortex (V1).
In their experimental methodology, they used moving light bars that made it possible to capture qualitative properties of receptive fields, such as the orientation selectivity in V1.
Later studies based on more refined stimuli, such as
white-noise-patterns, have then made it possible to reconstruct more detailed characterizations of biological visual receptive fields from multiple measurements of the same cell
(DeAngelis {\em et al.\/}\
\citeyear{DeAngOhzFre95-TINS,deAngAnz04-VisNeuroSci};
Ringach \citeyear{Rin01-JNeuroPhys,Rin04-JPhys};
Conway and Livingstone \citeyear{ConLiv06-JNeurSci};
Johnson {\em et al.\/}\ \citeyear{JohHawSha08-JNeuroSci};
Ghodrati {\em et al.\/} \citeyear{GhoKhaLeh17-ProNeurobiol};
De and Horwitz \citeyear{DeHor21-JNPhys}).
Summarizing these results in a qualitative manner,
it has been found that a majority of the receptive fields
in the retina and the LGN are rotationally symmetric over the spatial
domain and space-time separable over the spatio-temporal domain,
whereas simple cells in V1 have strong orientation preference over the
spatial domain, as well as that the simple cells are either space-time separable over the
spatio-temporal domain or tuned to particular motion directions in
joint space-time.

Learning-based schemes, which learn receptive fields from collections
of training data, have been formulated, trying to explain those types of
receptive fields found in biologically vision.
Rao and Ballard (\citeyear{RaoBal98-CompNeurSyst}) demonstrated how
localized oriented receptive fields could be obtained by learning a
translation-invariant code for natural images.
Olshausen and Field (\citeyear{OlsFie96-Nature,OlsFie97-VR}) proposed that
properties of receptive fields similar to biological receptive fields
could be obtained by learning a sparse code for natural images.
Simoncelli and Olshausen (\citeyear{SimOls01-AnnRevNeurSci}),
Geisler (\citeyear{Wil08-AnnRevPsychol}) and
Hyv{\"a}rinen {\em et al.\/} (\citeyear{HyvHurHoy09-NatImgStat})
argued that the properties of  neural
representations are determined by natural image statistics. 
L{\"o}rincz {\em et al.\/} (\citeyear{LoePalSzi12-PLOS-CB}) proposed
that early sensory processing can be modelled by sparse coding.
Poggio and Anselmi (\citeyear{PogAns16-book}) proposed to model
learning of invariant receptive fields by using group theory.
Singer {\em et al.\/} (\citeyear{SinTerWilSchKinHar18-Elife})
used the proxy task of predicting the relative future in
pre-recorded video sequences for training a deep network, and
demonstrated how that approach lead to receptive field shapes with good qualitative
similarities to biological receptive fields.
Deep neural network approaches for analysing and modelling non-linear
receptive fields of sensory neural responses have also been developed
(Keshishian {\em et al.\/} \citeyear{KesAkbKhaHerMehMes20-Elife}); see also
the more general discussions concerning such methodologies in
(Bae {\em et al.\/} \citeyear{BaeKimKim21-FrontSystNeuroSci}),
 (Bowers {\em et al.\/}\
 \citeyear{BowMalDujMonTsvBisPueAdoHumHeaEvaMitBly22-BehBrainSci}),
 (Heinke {\em et al.\/}\ \citeyear{HeiLeoLee22-VisRes}) and
(Wichmann and Geirhos \citeyear{WichGei23-AnnRevVisSci}) and the references therein.

Mathematically based computational models have also been formulated to reflect the shapes of
receptive fields found in biological vision.
Rodieck (\citeyear{Rod65-VisRes}) proposed to model circularly symmetric
receptive fields in the retina and the lateral geniculate nucleus
(LGN) by differences-of-Gaussians.
Marcelja (\citeyear{Mar80-JOSA}) as well as
Jones and Palmer (\citeyear{JonPal87a,JonPal87b}) proposed to model simple cells
by Gabor functions, motivated by their property of
minimizing the uncertainty relation; see also (Porat and Zeevi
\citeyear{PorZee88-PAMI}) for a more general proposal of using the
Gabor filter model for visual operations.
Riesenhuber and Poggio (\citeyear{RiePog99-Nature}) built on these
ideas, and used Gabor functions in a hierarchical model of object recognition.
Young and his co-workers
(\citeyear{You87-SV,YouLesMey01-SV,YouLes01-SV})
proposed to instead model simple cells by Gaussian
derivatives, with close relations to theoretical arguments in support
for the (regular) Gaussian derivative model stated by
Koenderink and van Doorn (\citeyear{Koe84,
  KoeDoo87-BC,KoeDoo92-PAMI}).
More detailed models of biological receptive fields based on the
associated (regular) Gaussian derivative framework have, in turn, been presented by
Lowe (\citeyear{Low00-BIO}),
May and Georgeson (\citeyear{MayGeo05-VisRes})
Hesse and Georgeson (\citeyear{HesGeo05-VisRes}),
Georgeson  {\em et al.\/}\ (\citeyear{GeoMayFreHes07-JVis}),
Wallis and Georgeson (\citeyear{WalGeo09-VisRes}),
Hansen and Neumann (\citeyear{HanNeu09-JVis}),
Wang and Spratling (\citeyear{WanSpra16-CognComp}) and
Pei {\em et al.\/}\ (\citeyear{PeiGaoHaoQiaAi16-NeurRegen}).

The generalized Gaussian derivative theory for visual receptive fields,
that we have built upon in this work, can
mathematically derive receptive field shapes directly from
symmetry properties of the environment in an axiomatic manner
(Lindeberg \citeyear{Lin10-JMIV,Lin13-BICY,Lin16-JMIV,Lin21-Heliyon})
regarding a
first layer of linear receptive fields, and can well model biological
receptive fields in the retina, the lateral geniculate nucleus and the
primary visual cortex. A conceptual  similarity
between this theoretical approach and the above learning-based approaches
is that the structural properties of the environment
will imply strong constraints on the statistics of natural
images, and thus the properties of the training data that the receptive
field shapes are learned from. Starting directly from the symmetry properties of the world,
thereby shortcircuits the need for learning receptive field shapes
from collections of training data, provided that the mathematical
analysis from the structural assumptions to the receptive field
shapes can be tractable.

\subsection{Extensions to non-linear visual receptive fields and artificial deep
  networks}

While the presented theory can be seen as theoretically rather
complete, as a model for the earliest layers of linear receptive
fields in an artificial vision system, or for the biological receptive
fields in the retina, the lateral geniculate nucleus and the primary visual cortex,
including their relations to the influence of geometric image
transformations, an interesting problem concerns how to extend the theory to higher
layers in the visual hierarchy, as well as to non-linear image and video
operations. Regarding the specific problem of achieving spatial scale
covariance in a non-linear hierarchy of visual receptive fields, a
general theoretical sufficiency result was presented in
(Lindeberg \citeyear{Lin20-JMIV}), which guarantees spatial scale covariance
for a hierarchical vision model based on a set of homogeneous
non-linear polynomial or rational combinations of scale-normalized Gaussian derivatives
coupled in cascade, including pointwise self-similar transformations
thereof. For a specific biomimetic implementation of this
general idea, in terms of an oriented quasi quadrature model that
reproduces some of the
known qualitative properties of complex cells, a scale-covariant hierarchical network
architecture was formulated, with close conceptual
similarities to the scattering network formulation of deep networks
proposed by Mallat and his co-workers (\citeyear{Mal16-RoySoc}).
Specifically, it was demonstrated
that, due to the scale invariant properties that arise from the resulting
provably scale-covariant network, it
was possible to perform predictions over spatial scales, to perform
training at one scale and testing at other scales, not spanned by the
training data (see Figures~15 and~16 in (Lindeberg \citeyear{Lin20-JMIV})).

More developed approaches to such
scale generalization based on scale-covariant and scale-invariant deep
networks were then presented in (Lindeberg \citeyear{Lin22-JMIV}) and
(Jansson and Lindeberg \citeyear{JanLin22-JMIV}). The approach in
(Lindeberg \citeyear{Lin22-JMIV}) is based on coupling linear
combinations of scale-normalized Gaussian derivatives in cascade, with
pointwise non-linearities
between, with close similarity to the previous work 
on using Gaussian derivative kernels as structured receptive
field models in deep networks  by
Jacobsen {\em et al.\/} (\citeyear{JacGemLouSme16-CVPR});
see also Pintea {\em et al.\/} (\citeyear{PinTomGoeLooGem21-TIP})
and Penaud {\em et al.\/} (\citeyear{PenVelAng22-ICIP}) for parallel work on using
Gaussian derivatives as primitive filters in deep networks. The approach in
(Jansson and Lindeberg \citeyear{JanLin22-JMIV}) is instead based on
building a deep network with multiple spatial scale channels, defined
by applying the same discrete deep network to multiple rescaled copies
of the input image, thus achieving scale-covariant and scale-invariant
properties in a dual manner, by performing multiple rescalings of the
input image, as opposed to applying multiple spatially rescaled spatial
receptive fields to the same input image, and leading to very good scale
generalization properties in experiments. It was also demonstrated how the resulting
scale-invariant multi-scale network was able to learn more efficiently from sparse
training data, compared to a single-scale network, in that
the multiple spatial scale channels could support each other in the
training phase, and make more efficient use of multi-scale training
data than a regular single-scale network.
Sangalli {\em et al.\/} (\citeyear{SanBluVelAng22-BMVC})
and Yang  {\em et al.\/} (\citeyear{YanDasMah23-arXiv}) have
performed closely related work on scale generalization based on
scale-covariant U-Nets.

More generally, Worrall and Welling (\citeyear{WorWel19-NeuroIPS}),
Sosnovik {\em et al.\/}
(\citeyear{SosSzmSme20-ICLR,SosMosSme21-BMVC,SosMosSme21-ICCV}),
Bekkers (\citeyear{Bek20-ICLR}) and
Zhu {\em et al.\/} (\citeyear{ZhuQiuCalSapChe22-JMLR}) have developed
scale-covariant or scale-equivariant deep network architectures, and
demonstrated that these lead to more robust results under variations
in the scale of image structures, compared to
non-covariant or non-equivariant counterparts.
Barisin {\em et al.\/} (\citeyear{BarSchRed23-arXiv}) have developed related
methods for handling multi-scale image structures and performing
scale generalization based on scale-invariant Riesz networks.

Currently, there is an active area of research to develop covariant or
equivariant deep networks, where we propose that it should be natural
to consider generalizations of the covariance properties under natural
image transformations treated for the receptive field models in this article.
We do also more generally propose to include more explicit treatments
of the influence of natural image transformations, such as covariance
and invariance properties under the classes of geometric image and
video transformations studied in this article,
in both the study as well as the computational modelling of
biological vision.

\section*{Acknowledgments}

This work was supported by the Swedish Research Council under
contracts 2018-03586 and 2022-02969.

\bibliographystyle{Frontiers-Harvard} 
\bibliography{bib/defs,bib/tlmac}


\section*{Figures}


\begin{figure}[hbt]
  \[
    \begin{CD}
       \hspace{-5mm}L_L(x_L;\; s, \Sigma_L)
       @>{\footnotesize\begin{array}{c}x_R=A x_L\\\Sigma_R=A \Sigma_L A^T \end{array}}>> L_R(x_R;\; s, \Sigma_R) \\
       \Big\uparrow\vcenter{\rlap{$\scriptstyle{{*g(x_L;\; s, \Sigma_L)}}$}} & &
       \Big\uparrow\vcenter{\rlap{$\scriptstyle{{*g(x_R;\; s, \Sigma_R)}}$}} \\
       f_L(x_L) @>{\footnotesize x_R= A x_L}>> f_R(x_R)
    \end{CD}
  \]
\caption{Commutative diagram for the spatial smoothing component in
  the purely spatial receptive field model (\ref{eq-spat-RF-model})
  under {\em spatial affine transformations\/} of the image domain. This
  commutative diagram, which should be read from the lower left corner to the
  upper right corner, means that irrespective of the input image
  $f_L(x_L)$ is first subject to a spatial affine transformation $x_R = A x_L$
  and then smoothed with an affine Gaussian kernel
  $g(x_R;\; s, \Sigma_R)$, or instead directly convolved
  with an affine Gaussian kernel $g(x_L;\; s, \Sigma_L)$ and then
  subject to a similar affine transformation, we do then get the same
  result, provided that the spatial covariance matrices are related
  according to $\Sigma_R = A \Sigma A^T$ (and assuming that the spatial
  scale parameters for the two domain $s_R = s_L = s$ are the same).}
\label{fig-comm-diag-aff-transf-spat}
\end{figure}

\begin{figure}[hbt]
  \[
    \begin{CD}
       L_L(x_L;\; s_L, \Sigma)
       @>{\footnotesize\begin{array}{c}x_R=S_x x_L\\s_R=S_x^2
                                           s_L\end{array}}>> L_R(x_R;\; s_R, \Sigma) \\
       \Big\uparrow\vcenter{\rlap{$\scriptstyle{{*g(x_L;\; s_L,\Sigma)}}$}} & &
       \Big\uparrow\vcenter{\rlap{$\scriptstyle{{*g(x_R;\; s_R, \Sigma)}}$}} \\
       f_L(x_L) @>{\footnotesize x_R = S_x x_R}>> f_R(x_R)
    \end{CD}
  \]
 \caption{Commutative diagram for the spatial smoothing component in
  the purely spatial receptive field model (\ref{eq-spat-RF-model})
  under {\em spatial scaling transformations\/} of the image domain. This
  commutative diagram, which should be read from the lower left corner to the
  upper right corner, means that irrespective of the input image
  $f_L(x_L)$ is first subject to spatial scaling transformation $x_R = S_x x_L$
  and then smoothed with an affine Gaussian kernel
  $g(x_R;\; s_R, \Sigma)$, or instead directly convolved
  with an affine Gaussian kernel $g(x_L;\; s_L, \Sigma)$ and then
  subject to a similar affine transformation, we do then get the same
  result, provided that the spatial scale parameters are related
  according to $s_R = S_x^2 s_L$ (and assuming that the spatial
  covariance matrices for the two image domains are the same $\Sigma_R = \Sigma_L = \Sigma$).}
\label{fig-comm-diag-spat-scaling}
\end{figure}

\begin{figure}[hbt]
  \[
    \begin{CD}
       \hspace{-7mm}L(x, t;\; s, \tau, v, \Sigma)
       @>{\footnotesize\begin{array}{c} x' = x + u t  \\ v' = v +
                         u \end{array}}>> L'(x', t;\; s, \tau, v', \Sigma) \\
       \Big\uparrow\vcenter{\rlap{$\scriptstyle{{*T(x, t;\; s,
               \tau, v, \Sigma)}}$}} & &
       \Big\uparrow\vcenter{\rlap{$\scriptstyle{{*T(x', t;\; s,
               \tau, v', \Sigma)}}$}} \\
       f(x, t) @>{\footnotesize x' = x + u t}>> f'(x', t)
    \end{CD}
  \]
\caption{Commutative diagram for the spatio-temporal smoothing
  component in the joint spatio-temporal receptive field
  model (\ref{eq-spat-temp-RF-model-der-norm-caus}) under
  {\em Galilean transformations\/} over a spatio-temporal video domain. This
  commutative diagram, which should be read from the lower left corner to the
  upper right corner, means that irrespective of the input video 
  $f(x, t)$ is first subject to Galilean transformation $x' = x' + u$
  and then smoothed with a spatio-temporal kernel kernel
  $T(x', t;\; s, \tau, v', \Sigma)$, or instead directly convolved
  with the spatio-temporal smoothing kernel
  $T(x, t;\; s, \tau, v, \Sigma)$ and then
  subject to a similar Galilean transformation, we do then get the same
  result, provided that the velocity parameters of the spatio-temporal
  smoothing kernels are related
  according to $v' = v + u$ (and assuming that the other parameters in the
  spatio-temporal receptive field models are the same).}
\label{fig-comm-diag-gal-transf}
\end{figure}

\begin{figure}[hbt]
  \[
    \begin{CD}
       \hspace{-7mm}L(x, t;\; s, \tau, v, \Sigma)
       @>{\footnotesize\begin{array}{c} t' = S_t t  \\ \tau' = S_t^2 \tau \\
                         v' = v/S_t \end{array}}>> L'(x, t';\; s, \tau', v', \Sigma) \\
       \Big\uparrow\vcenter{\rlap{$\scriptstyle{{*T(x, t;\; s,
               \tau, v, \Sigma)}}$}} & &
       \Big\uparrow\vcenter{\rlap{$\scriptstyle{{*T(x, t';\; s,
               \tau', v', \Sigma)}}$}} \\
       f(x, t) @>{\footnotesize t' = S_t t}>> f'(x, t')
    \end{CD}
 \]
\caption{Commutative diagram for the spatio-temporal smoothing
  component in the joint spatio-temporal receptive field
  model (\ref{eq-spat-temp-RF-model-der-norm-caus}) under
  {\em temporal scaling transformations\/} over a spatio-temporal video domain. This
  commutative diagram, which should be read from the lower left corner to the
  upper right corner, means that irrespective of the input video 
  $f(x, t)$ is first subject to a temporal scaling transformation $t' = S_t t$
  and then smoothed with a spatio-temporal kernel kernel
  $T(x, t';\; s, \tau', v', \Sigma)$, or instead directly convolved
  with the spatio-temporal smoothing kernel
  $T(x, t;\; s, \tau, v, \Sigma)$ and then
  subject to a similar temporal scaling transformation, we do then get the same
  result, provided that the temporal scale parameters as well as the
  velocity parameters of the spatio-temporal
  smoothing kernels are matched
  according to $\tau' = S_t^2 t$ and $v' = v/S_t$
  (and assuming that the other parameters in the
  spatio-temporal receptive field models are the same).}
\label{fig-comm-diag-temp-sc-transf}
\end{figure}

\begin{figure}[hbt]
  \[
    \begin{CD}
       \hspace{-7mm}L_L(x_L, t;\; s, \tau, v_L, \Sigma_L)
       @>{\footnotesize\begin{array}{c} x_R = A x_L  \\ 
                        \Sigma_R=A  \Sigma_L A^T \\ v_R = A v_L
                         \end{array}}>> L_R(x_R, t;\; s, \tau, v_R, \Sigma_R) \\
       \Big\uparrow\vcenter{\rlap{$\scriptstyle{{*T(x_L, t;\; s,
               \tau, v_L, \Sigma_L)}}$}} & &
       \Big\uparrow\vcenter{\rlap{$\scriptstyle{{*T(x_R, t;\; s,
               \tau, v_R, \Sigma_R)}}$}} \\
       f_L(x_L, t) @>{\footnotesize x_R = A x_L}>> f_R(x_R, t)
    \end{CD}
 \]
\caption{Commutative diagram for the spatio-temporal smoothing
  component in the joint spatio-temporal receptive field
  model (\ref{eq-spat-temp-RF-model-der-norm-caus}) under
  {\em spatial affine transformations\/} over a spatio-temporal video domain. This
  commutative diagram, which should be read from the lower left corner to the
  upper right corner, means that irrespective of the input video 
  $f(x_L, t)$ is first subject to a spatial affine transformation $x_R = A  x_L$
  and then smoothed with a spatio-temporal kernel kernel
  $T(x_R, t;\; s, \tau, v_R, \Sigma_R)$, or instead directly convolved
  with the spatio-temporal smoothing kernel
  $T(x_L, t;\; s, \tau, v_L, \Sigma_L)$ and then
  subject to a similar temporal scaling transformation, we do then get the same
  result, provided that the spatial covariance matrices as well as the
  velocity parameters of the spatio-temporal
  smoothing kernels are matched
  according to $\Sigma_R = A  \Sigma_L A^T $ and $v_R = A v_L$
  (and assuming that the other parameters in the
  spatio-temporal receptive field models are the same).}
\label{fig-comm-diag-aff-transf-spat-temp}
\end{figure}

\begin{figure}[hbt]
  \[
    \begin{CD}
       \hspace{-7mm}L(x_L, t;\; s_L, \tau, v_L, \Sigma)
       @>{\footnotesize\begin{array}{c} x_R = S x_L  \\ s_R = S_x^2 s_L \\ v_R = S_x v_L
                         \end{array}}>> L_R(x_R, t;\; s_R, \tau, v_R, \Sigma) \\
       \Big\uparrow\vcenter{\rlap{$\scriptstyle{{*T(x_L, t;\; s_L,
               \tau, v_L, \Sigma)}}$}} & &
       \Big\uparrow\vcenter{\rlap{$\scriptstyle{{*T(x_R, t;\; s_R,
               \tau;\; v_R, \Sigma)}}$}} \\
       f(x_L, t) @>{\footnotesize x_R = S_x x_L}>> f_R(x_R, t)
    \end{CD}
 \]
\caption{Commutative diagram for the spatio-temporal smoothing
  component in the joint spatio-temporal receptive field
  model (\ref{eq-spat-temp-RF-model-der-norm-caus}) under
  {\em spatial scaling transformations\/} over a spatio-temporal video domain. This
  commutative diagram, which should be read from the lower left corner to the
  upper right corner, means that irrespective of the input video 
  $f(x_L, t)$ is first subject to a spatial scaling transformation $x_R = S_x  x_L$
  and then smoothed with a spatio-temporal kernel kernel
  $T(x_R, t;\; s_R, \tau, v_R, \Sigma)$, or instead directly convolved
  with the spatio-temporal smoothing kernel
  $T(x_L, t;\; s_L, \tau, v_L, \Sigma)$ and then
  subject to a similar temporal scaling transformation, we do then get the same
  result, provided that the spatial scale parameters as well as the
  velocity parameters of the spatio-temporal
  smoothing kernels are matched
  according to $s_R = S_x^2 s_L$ and $v_R = S_x v_L$
  (and assuming that the other parameters in the
  spatio-temporal receptive field models are the same).}
\label{fig-comm-diag-sc-transf-spat-temp}
\end{figure}

\begin{figure}[!b]
  \begin{center}
    \begin{tabular}{c}
     \includegraphics[width=0.4\textwidth]{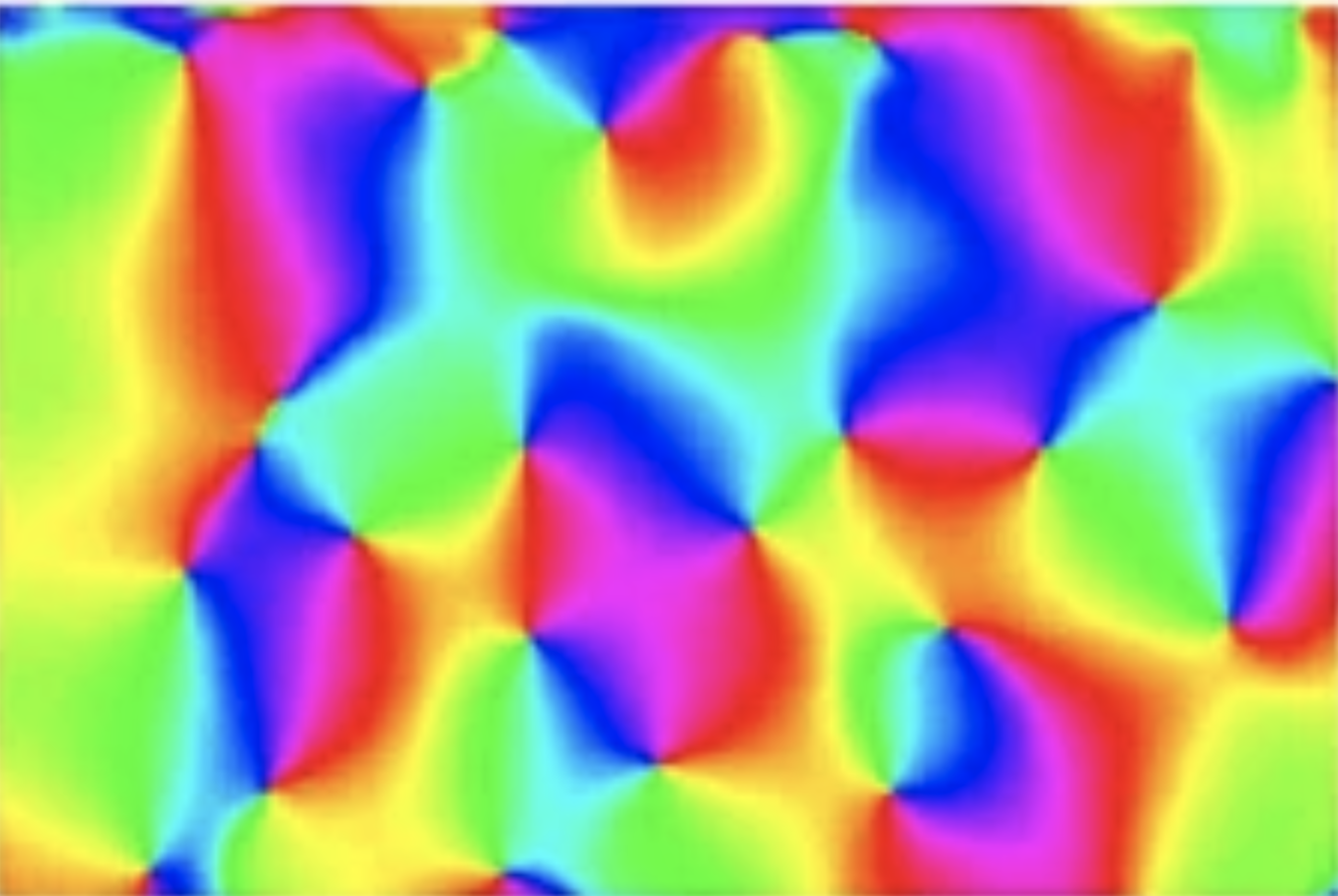}
    \end{tabular} 
    \caption{Orientation map in the primary visual cortex (in cat area 17), 
      as recorded by Koch {\em et al.\/} (\citeyear{KocJinAloZai16-NatComm}) (OpenAccess),
     which by a colour coding
    shows how the preferred spatial orientations for the visual
    neurons vary spatially over the cortex. In terms of the theory
    presented in this article, regarding
    connections between the influence of natural image transformation
    on image data and the shapes of the visual receptive fields of the
    neurons that process the visual information,
    these results can be taken as support that the
    population of visual receptive fields in the visual cortex perform
    an expansion over the group of spatial rotations. Let us take into
    further account that the receptive fields of the neurons near the
    pinwheels are reported to be comparably isotropic, whereas the
    receptive fields of the neurons further away from the pinwheels
    are more anisotropic. Then, we can, from the prediction of an expansion
    over affine covariance matrices of directional derivatives of
    affine Gaussian kernels, proposed as a model for the spatial
    component of simple cells, raise the question if the population of
    receptive fields can also be regarded as spanning some larger
    subset of the affine group than mere rotations. If the receptive
    fields of the neurons span a larger part of the affine group,
    involving non-uniform scaling transformations that would correspond to
    spatial receptive fields with different ratios between the
    characteristic lengths of the receptive fields in the directions
    of the orientation of the receptive field and its orthogonal
    direction, then
  they would have better ability to derive properties, such as the
  surface orientation and the surface shape of objects in the world,
  compared to a population of receptive fields that does not perform
  such an expansion.}
  \label{fig-orient-maps}
  \end{center}
\end{figure}


\end{document}